\documentclass[preprint, superscriptaddress,preprintnumbers,amsmath,amssymb,pra]{revtex4}
\usepackage{graphicx}

\usepackage{graphicx}

\usepackage{color}

\newcommand{\ve}{\varepsilon}
\newcommand{\tq}{\tilde{q}}
\newcommand{\tql}{\tilde{q}_l}
\newcommand{\add}[1]{{\color{magenta} #1}}

\begin{document}

\newcommand{\be}{\begin{equation}}
\newcommand{\ee}{\end{equation}}

\thispagestyle{empty}

\title{Thermal effect in the Casimir force for graphene and graphene-coated
substrates: Impact of nonzero mass gap and chemical potential}

\author{
G.~Bimonte}
\affiliation{Dipartimento di Fisica E. Pancini, Universit\`{a} di Napoli
Federico II, Complesso Universitario MSA, Via Cintia, I-80126 Napoli, Italy}
\affiliation{INFN Sezione di Napoli, I-80126, Napoli, Italy}

\author{
G.~L.~Klimchitskaya}
\affiliation{Central Astronomical Observatory at Pulkovo of the
Russian Academy of Sciences, Saint Petersburg,
196140, Russia}
\affiliation{Institute of Physics, Nanotechnology and
Telecommunications, Peter the Great Saint Petersburg
Polytechnic University, Saint Petersburg, 195251, Russia}

\author{
V.~M.~Mostepanenko}
\affiliation{Central Astronomical Observatory at Pulkovo of the
Russian Academy of Sciences, Saint Petersburg,
196140, Russia}
\affiliation{Institute of Physics, Nanotechnology and
Telecommunications, Peter the Great Saint Petersburg
Polytechnic University, Saint Petersburg, 195251, Russia}
\affiliation{Kazan Federal University, Kazan, 420008, Russia}

\begin{abstract}
The rigorous finite-temperature QED  formalism of the polarization tensor
is  used to study the combined effect of  nonzero mass gap $m$ and chemical
potential $\mu$ on the Casimir force and its thermal correction in the
experimentally relevant configuration of a Au sphere interacting with a real
graphene sheet or  with graphene-coated dielectric substrates made of different
materials. It is shown that for  both a free-standing graphene sheet and for
graphene-coated substrates the magnitude of
the Casimir force decreases as $m$  is increased, while it increases as $\mu$ is
increased, indicating that these parameters act in
opposite directions. According to our results, the impact of
$m$ and/or $\mu$ on the Casimir force for graphene-coated plates is much
smaller than for a free-standing graphene sheet. Furthermore,
computations show that  the  Casimir force is much stronger  for  graphene-coated substrates than for a free-standing
graphene sample, but the thermal correction and its fractional weight in the total
force are smaller in the former case. These results are applied to a differential setup that was recently proposed to observe the giant thermal effect in the  Casimir force for graphene. We show that this experiment remains feasible even after taking into account the influence of the nonzero mass-gap and chemical potential of real graphene samples. Possible further applications of the obtained results
are discussed.
\end{abstract}

\maketitle

\section{INTRODUCTION}

Recent trends are toward increased use of carbon nanostructures,
such as buckyballs, nanotubes, nanowires and graphene, in a variety of
applications to microelectronics \cite{1,2}.
Graphene occupies a prominent place among these new materials since
its investigation led to many important experimental and theoretical
discoveries \cite{2,3}. Specifically, several fascinating effects
have been found for graphene interacting with magnetic and electric
fields \cite{4,5,6,7,8,9}. A consensus on the value of the universal
electrical conductivity of graphene $e^2/(4\hbar)$ expressed {in terms of} the
electron charge $e$ and Planck constant $\hbar$ has been achieved
\cite{10,11,12,13,14,15,16,17,18}.
The reflectivity properties of graphene  and graphene-coated substrates
{have been} investigated as functions of frequency and temperature
{revealing} some unusual properties \cite{13,19,20,21,22,23,24,24a}.

The Casimir effect in graphene systems has attracted widespread attention
shortly after the advent of graphene. The Casimir force arises between
two closely spaced material surfaces {as a result of}  zero-point and thermal fluctuations
of the electromagnetic field \cite{25}. In the framework of the Lifshitz
theory \cite{25,26} the Casimir force between two dissimilar 3D-materials
at any temperature $T$ is routinely represented as a functional of their
reflection coefficients {evaluated} at the pure imaginary Matsubara
frequencies. These coefficients are usually expressed {in terms of} the frequency-dependent
dielectric permittivities of both materials. {Since}
graphene is a one-atom-thick hexagonal sheet of carbon atoms, {its response to external electromagnetic fields}
is, strictly speaking, nonlocal and
cannot be
described by {a} dielectric permittivity
depending only on frequency.
That is why { early applications of} the
Lifshitz theory {to} graphene {adopted a}
hydrodynamic {approach in which graphene was modelled} as a two-dimensional electronic fluid characterized by some
typical wave number \cite{27,28,29}. At a later time, the hydrodynamic model
was used for {a} theoretical description of the Casimir and Casimir-Polder
interactions with different carbon nanostructures \cite{30,31,32,33}.
{Unfortunately, it turned out} \cite{34}  that theoretical predictions obtained using
the hydrodynamic model are excluded by measurements of the gradient of
Casimir force between a Au-coated sphere and a graphene-coated SiO${}_2$
film deposited on a Si plate \cite{35}.

The literature on the Casimir effect in graphene systems is quite extensive.
Currently most of the used calculation approaches are {based on} the
Dirac model for graphene. According to this model, at energies below a few
electron volts the quasiparticles in graphene are massless{,} and satisfy {a}
linear dispersion relation {in which the speed of light} $c$ is replaced with the Fermi velocity
$v_{\rm F}\approx c/300$ \cite{2,3,36,37a}. Calculations of the Casimir
(Casimir-Polder) force between two graphene sheets, {a} graphene sheet and
a 3D-material plate, graphene-coated substrates, and an atom and a graphene
sheet have been performed using the density-density correlation functions in
the random phase approximation, by modelling the conductivity of graphene as
a combination of Lorentz-type oscillators, and {within} the Kubo formalism
\cite{37,38,39,40,41,42a,42,43,44,45,46,47a,47b,47c,47,48a}.
Some of the results obtained {in these ways} were reviewed in Ref.~\cite{48}.
The most impressive result for the Casimir force was obtained in Ref.~\cite{38},
where it was found that the thermal correction to the force becomes dominant at
much shorter separations, as compared to the case of 3D interacting bodies.

The fundamental approach for obtaining the response function for a material
body to electromagnetic field consists in {the} calculation of its polarization
tensor \cite{49,50}. For a graphene sheet described by the Dirac model the
polarization tensor in (2+1)-dimensional space-time (and, thus, the in-plane
and out-of-plane nonlocal dielectric permittivities and conductivities of
graphene) can be {worked out exactly}. This was done at $T=0$ in Ref.~\cite{51}
and at nonzero $T$ in Ref.~\cite{52} for graphene {in the cases of} both {vanishing} and
{nonvanishing}  quasiparticle {mass} $m$ {corresponding to a} gap
$\Delta=2mc^2$ (it should be noted that for  $T\neq 0$ the {expression for the} polarization tensor of Ref.~\cite{52}
 is valid only at the
discrete imaginary Matsubara frequencies {occurring} in the Lifshitz {formula}).
The reflection coefficients {of} graphene have been expressed via the components
of the polarization tensor and used to calculate the Casimir force between
a graphene sheet and an ideal metal plane \cite{51,52}.
At a later time, the results of Refs.~\cite{51,52} have been used to calculate
the thermal Casimir and Casimir-Polder force in many physical systems including
two graphene sheets, {a} graphene {sheet} and {a} plate made of various real materials,
graphene-coated substrates, atom and graphene or graphene-coated substrates etc.
\cite{53,54,55,56,57,58,59,60,61}.
In so doing, the role of {a nonvanishing}  mass gap of graphene was investigated, and
the {existence} of a giant thermal effect at short separations \cite{38} was
confirmed. It was shown \cite{62} that the formalism of {the} polarization tensor
is in fact equivalent to the formalism of the density-density correlation
functions, but the former is somewhat {preferable} because the latter
quantities {have not been known} precisely. The theoretical predictions for
the gradient of the Casimir force computed using the polarization tensor have
been shown to be in a very good agreement \cite{63} with the measurement data of
Ref.~\cite{35}.

{A different} representation for the polarization tensor of graphene allowing {this time for} an analytic
continuation to the entire plane of complex frequencies was derived in Ref.~\cite{22}
for both zero and nonzero mass gap. {The novel representation} was applied {to the} investigation of the giant
thermal effect \cite{64,64a} and to test the validity of the Nernst heat theorem for
the Casimir entropy in graphene systems \cite{65}. After an analytic continuation
to the real frequency axis, the polarization tensor of Ref.~\cite{22} has been
used to describe the electrical conductivity and reflectivity properties of graphene
\cite{17,18,22,23,24,24a}. In Ref.~\cite{66} this tensor was further generalized
to the case of doped graphene with nonzero chemical potential.
It was shown that for doped but gapless graphene characterized by nonzero chemical
potential the thermal Casimir force between a graphene sheet and an ideal metal plane
can be enhanced up to 60\% as compared to the case of a pristine (undoped) graphene.

In this paper, we investigate the thermal Casimir force in the experimentally
relevant configuration of a Au-coated sphere above a real graphene sheet
characterized by nonzero values {of}  the mass gap $m$ {and/or the} chemical potential $\mu$. The case of a dielectric plate coated with {a} real graphene
sheet is also considered. Using the polarization tensor of graphene in the form
of Refs.~\cite{22,66}, we perform calculations of both the Casimir force and {its
room-temperature}
thermal correction   for a free-standing graphene
characterized by nonzero values of $m$ and $\mu$, as well as for graphene deposited on
SiO${}_2$ and Si plates. It is shown that with fixed $\mu$ and increasing $m$
the magnitude of the Casimir force decreases. By contrast, with fixed $m$
and increasing $\mu$  the magnitude of the Casimir force increases.
This means that for real graphene  (both free-standing and deposited on a substrate)
the impacts of nonzero mass gap and chemical potential on the Casimir force
partially compensate each other. Another important result found is that the impacts
of both nonzero $m$ and $\mu$ on the Casimir force for graphene-coated substrates
 are much smaller than {the corresponding effects} for a free-standing graphene.
Qualitatively, all the above results are quite expected and
{have a} simple
physical explanation.
It is interesting also that
 the thermal correction to the Casimir force is a nonmonotonous function of both
 $m$ and $\mu$.

 We also investigate the impact of nonzero $m$ and $\mu$ in the recently proposed
 differential measurement scheme \cite{67} which allows for a clear observation
 of the giant thermal effect for the Casimir force in graphene systems at short
 separations. For this purpose the differences among the Casimir forces between
 a Au-coated sphere and the two halves of a Si plate, one uncoated and the other
 coated with  graphene characterized by nonzero $m$ and $\mu$, are calculated
 at both room and zero temperature. It is shown that
 {the possible presence of nonvanishing}  $m$ and
 $\mu$  does not prevent a clear observation of the giant thermal
 effect for graphene in the proposed experiment at separation distances exceeding 220\,nm.

 The paper is organized as follows. In Sec.~II we present the general formalism
 describing the Casimir force between {a} metallic sphere and {a} real graphene or
 graphene-coated plate in terms of the polarization tensor with nonzero $m$ and $\mu$.
 In Sec.~III the role of nonzero $m$ and $\mu$ is {investigated} for the case of
a free-standing graphene sheet. Section~IV contains the computational results
 demonstrating a suppressed impact of nonzero $m$ and $\mu$ in the case of a graphene
sheet deposited on dielectric plate made  either {of} silica or silicon.
Section~V investigates the influence of nonzero $m$ and $\mu$ in the differential
measurement scheme, which was proposed to measure the thermal effect in graphene
systems. In Sec.~VI the reader will find our conclusions and discussion.

\section{General formalism for metallic sphere interacting with graphene
or graphene-coated substrate}

We consider a Au-coated sphere of radius $R$ spaced at a height $a$ above a dielectric
plate coated {by}   a real graphene {sheet} with nonzero quasiparticle mass $m$ and chemical
potential $\mu$. {In practice a} Au coating {with a thickness larger} than a few tens of nanometers  allows
to consider {the} sphere as  all-gold in calculations of the Casimir force \cite{25}.
The plate is assumed to be of sufficient thickness to consider it as a semispace.
The Casimir force between a sphere and a graphene-coated plate at temperature $T$ in
thermal equilibrium with {the} environment can be expressed by using the Lifshitz
formula and the proximity force approximation \cite{25,26}
\begin{equation}
F(a,T)=k_BTR\sum_{l=0}^{\infty}{\vphantom{\sum}}^{\prime}
\int_{0}^{\infty}k_{\bot}\,dk_{\bot}
\sum_{\alpha}\ln\left[1-
r_{\alpha}^{(1)}(i\xi_l,k_{\bot})R_{\alpha}^{(n)}(i\xi_l,k_{\bot})
e^{-2q_la}\right].
\label{eq1}
\end{equation}
\noindent
Here, $k_B$ is the Boltzmann constant, the prime in the first summation sign means
that the term with $l=0$ is taken with weight 1/2, $k_{\bot}$ is the magnitude
of the in-plane wave vector, $\xi_l=2\pi k_BTl/\hbar$ with $l=0,\,1,\,2,\,\ldots$
are the Matsubara frequencies, and $q_l=\sqrt{k_{\bot}^2+\xi_l^2/c^2}$.
The summation in $\alpha$ is over two independent polarizations of the electromagnetic
field, transverse magnetic ($\alpha={\rm TM}$) and transverse electric ($\alpha={\rm TE}$).

The reflection coefficients $r_{\alpha}$ on the boundary between Au and vacuum are
given by \cite{25}
\begin{eqnarray}
&&
r_{\rm TM}^{(1)}(i\xi_l,k_{\bot})=
\frac{\varepsilon_l^{(1)}q_l-k_l^{(1)}}{\varepsilon_l^{(1)}q_l+k_l^{(1)}},
\nonumber \\
&&
r_{\rm TE}^{(1)}(i\xi_l,k_{\bot})=
\frac{q_l-k_l^{(1)}}{q_l+k_l^{(1)}},
\label{eq2}
\end{eqnarray}
\noindent
where $\varepsilon_l^{(1)}\equiv \varepsilon^{(1)}(i\xi_l)$ is the dielectric
permittivity of Au calculated at the pure imaginary Matsubara frequencies, and
$k_l^{(1)}=\sqrt{k_{\bot}^2+\varepsilon_l^{(1)}\xi_l^2/c^2}$.
The reflection coefficients $R_{\alpha}^{(n)}$ on the boundary between {vacuum and the}
graphene-coated plate made of a {dielectric} material {(denoted by the superscript $n$)} take the form
\cite{61,63,64}
\begin{eqnarray}
&&
R_{\rm TM}^{(n)}(i\xi_l,k_{\bot})=
\frac{\hbar k_{\bot}^2(\varepsilon_l^{(n)}q_l-k_l^{(n)})+
q_lk_l^{(n)}\Pi_{00,l}}{\hbar k_{\bot}^2(\varepsilon_l^{({n})}q_l+k_l^{({n})})+
q_lk_l^{(n)}\Pi_{00,l}},
\nonumber \\
&&
R_{\rm TE}^{(n)}(i\xi_l,k_{\bot})=
\frac{\hbar k_{\bot}^2(q_l-k_l^{(n)})-\Pi_l}{\hbar k_{\bot}^2
(q_l+k_l^{(n)})+\Pi_l},
\label{eq3}
\end{eqnarray}
\noindent
where $\varepsilon_l^{(n)}\equiv \varepsilon^{(n)}(i\xi_l)$,
$n=1,\>2$ are the dielectric
permittivities of {the} two plate materials and
$k_l^{(n)}=\sqrt{k_{\bot}^2+\varepsilon_l^{(n)}\xi_l^2/c^2}$.
The quantities
$\Pi_{\beta\gamma,l}\equiv \Pi_{\beta\gamma}(i\xi_l,k_{\bot},T,m,\mu)$\
with $\beta,\,\gamma=0,\,1,\,2$ are the components of the polarization tensor
of graphene in (2+1)-dimensional space-time and  $\Pi_l$ {is}
defined {as}
\begin{equation}
\Pi_l=k_{\bot}^2\Pi_{{\rm tr},l}-q_l^2\Pi_{00,l}.
\label{eq4}
\end{equation}
\noindent
Here, $\Pi_{\rm tr}=\Pi_{\beta}^{\,\beta}$ is the trace of the polarization
tensor.

If the sphere interacts with {a} free-standing graphene sheet, one has
$\varepsilon_l^{(n)}=1$, $k_l^{(n)}=q_l$ and the reflection coefficients
(\ref{eq3}) transform to \cite{62,63}
\begin{eqnarray}
&&
R_{\rm TM}(i\xi_l,k_{\bot})=
\frac{q_l\Pi_{00,l}}{q_l\Pi_{00,l}+2\hbar k_{\bot}^2},
\nonumber \\
&&
R_{\rm TE}(i\xi_l,k_{\bot})=-
\frac{\Pi_l}{\Pi_l+2\hbar k_{\bot}^2q_l}.
\label{eq5}
\end{eqnarray}
\noindent
Note that the proximity force approximation used in the derivation of
Eq.~(\ref{eq1}) is valid under the condition $a\ll R$. Direct calculations
show that the relative correction to the PFA result (\ref{eq1}) is smaller
than $a/R$ \cite{68,69,70,71,71a,71b}.

Here we use the explicit expressions for the quantities $\Pi_{00,l}$ and
$\Pi_l$ in the case of graphene with nonzero $m$ and $\mu$ which allow
analytic continuation to the entire plane of complex frequencies.
It is convenient to present them as sums of two contributions \cite{66}
\begin{eqnarray}
&&
\Pi_{00}(i\xi_l,k_{\bot},T,m,\mu)=\Pi_{00}^{(0)}(i\xi_l,k_{\bot},m)
+\Pi_{00}^{(1)}(i\xi_l,k_{\bot},T,m,\mu),
\nonumber \\
&&
\Pi(i\xi_l,k_{\bot},T,m,\mu)=\Pi^{(0)}(i\xi_l,k_{\bot},m)
+\Pi^{(1)}(i\xi_l,k_{\bot},T,m,\mu).
\label{eq6}
\end{eqnarray}
\noindent
The first terms on the right-hand sides of Eq.~(\ref{eq6}), $\Pi_{00}^{(0)}$ and
$\Pi^{(0)}$, are the contributions to the polarization tensor describing undoped
($\mu=0$) graphene with  nonzero mass gap at zero temperature
calculated at the imaginary Matsubara frequencies. They were
obtained in Ref.~\cite{51} and can be equivalently presented in the form
\begin{eqnarray}
&&
\Pi_{00,l}^{(0)}=\frac{\alpha \hbar k_{\bot}^2}{\tql}\,\Psi\!\left(
\frac{2mc}{\hbar\tql}\right),
\nonumber \\
&&
\Pi_{l}^{(0)}={\alpha \hbar k_{\bot}^2}{\tql}\,\Psi\!\left(
\frac{2mc}{\hbar\tql}\right),
\label{eq7}
\end{eqnarray}
\noindent
where
\begin{equation}
\Psi(x)=2\left[x+(1-x^2)\arctan\frac{1}{x}\right],
\qquad
\tql=\sqrt{\frac{v_F^2}{c^2}k_{\bot}^2+\frac{\xi_l^2}{c^2}},
\label{eq8}
\end{equation}
\noindent
and $\alpha=e^2/(\hbar c)$ is the fine structure constant.

The second terms on the right-hand sides of Eq.~(\ref{eq6}) take into
account both the thermal effect and the dependence of the polarization
tensor on the chemical potential. For doped graphene the latter may
remain {different from zero} in the limiting case of
{vanishing} temperature. The resulting
$\mu$-dependent contributions to the polarization tensor depend also on $m$
(see below). The explicit expressions for the
second terms on the right-hand sides of Eq.~(\ref{eq6}),
$\Pi_{00}^{(1)}$ and $\Pi^{(1)}$, were derived in Ref.~\cite{66}.
They can be equivalently presented as
\begin{eqnarray}
&&
\Pi_{00,l}^{(1)}=\frac{4\alpha\hbar c^2\tql}{v_F^2}
\int_{D_l}^{\infty}du \left(
\frac{1}{e^{B_lu+\frac{\mu}{k_BT}}+1}+
\frac{1}{e^{B_lu-\frac{\mu}{k_BT}}+1}\right)
\nonumber \\
&&
~\times\left[1-{\rm Re}\frac{1-u^2+2i\frac{\xi_l}{c\tql}u}{\left(1-u^2
+2i\frac{\xi_l}{c\tql}u+\frac{v_F^2k_{\bot}^2}{c^2\tql^2}D_l^2\right)^{1/2}}
\right],
\nonumber \\
&&
\Pi_{l}^{(1)}=-\frac{4\alpha\hbar \tql\xi_l^2}{v_F^2}
\int_{D_l}^{\infty}du \left(
\frac{1}{e^{B_lu+\frac{\mu}{k_BT}}+1}+
\frac{1}{e^{B_lu-\frac{\mu}{k_BT}}+1}\right)
\nonumber \\
&&
~\times\left[1-{\rm Re}\frac{1-\tql^2\frac{c^2}{\xi_l^2}u^2+2i\frac{c\tql}{\xi_l}u+
\frac{v_F^2k_{\bot}^2}{\xi_l^2}D_l^2}{\left(1-u^2
+2i\frac{\xi_l}{c\tql}u+\frac{v_F^2k_{\bot}^2}{c^2\tql^2}D_l^2\right)^{1/2}}
\right],
\label{eq9}
\end{eqnarray}
\noindent
where
\begin{equation}
D_l=\frac{2mc}{\hbar\tql}, \quad
B_l=\frac{\hbar c\tql}{2k_BT}.
\label{eq10}
\end{equation}

Note that in the framework of quantum field theory at nonzero temperature
the chemical potential is introduced by the substitution \cite{72}
\begin{equation}
\frac{1}{\exp(B_lu)+1}\to \frac{1}{2}
\left(
\frac{1}{e^{B_lu+\frac{\mu}{k_BT}}+1}+
\frac{1}{e^{B_lu-\frac{\mu}{k_BT}}+1}\right).
\label{eq11}
\end{equation}
\noindent
Using this equation, the results (\ref{eq9}) follow also from the
respective equations of Ref.~\cite{22} obtained for the case $m\neq 0$, $\mu=0$.

It is convenient to consider separately the zero-frequency contribution to
Eq.~(\ref{eq1}), $l=0$, and {the} contributions {of} nonzero Matsubara frequencies
with $l\geq 1$.  Equations~(\ref{eq6}), (\ref{eq7}), and (\ref{eq9})
for the  polarization tensor at $l=0$ take the form
\begin{eqnarray}
&&
\Pi_{00,0}=\alpha\hbar c\frac{k_{\bot}}{v_F}\,
\Psi\!\left(\frac{2mc^2}{\hbar v_Fk_{\bot}}\right)+
\frac{8\alpha k_BTc}{v_F^2}\ln\left[\left(e^{\frac{\mu}{k_BT}}+
e^{-\frac{mc^2}{k_BT}}\right)\left(e^{-\frac{\mu}{k_BT}}+
e^{-\frac{mc^2}{k_BT}}\right)\right]
\nonumber \\
&&
~
-\frac{4\alpha\hbar c k_{\bot}}{v_F}
\int_{D_0}^{\sqrt{1+D_0^2}}du
\left(
\frac{1}{e^{B_lu+\frac{\mu}{k_BT}}+1}+
\frac{1}{e^{B_lu-\frac{\mu}{k_BT}}+1}\right)
\frac{1-u^2}{\sqrt{1-u^2+D_0^2}},
\nonumber \\
&&
\Pi_{0}=\alpha\hbar \frac{v_Fk_{\bot}^3}{c}\,
\Psi\!\left(\frac{2mc^2}{\hbar v_Fk_{\bot}}\right)
\label{eq12} \\
&&
~
+4\alpha\hbar\frac{ v_Fk_{\bot}^3}{c}
\int_{D_0}^{\sqrt{1+D_0^2}}du
\left(
\frac{1}{e^{B_lu+\frac{\mu}{k_BT}}+1}+
\frac{1}{e^{B_lu-\frac{\mu}{k_BT}}+1}\right)
\frac{-u^2+D_0^2}{\sqrt{1-u^2+D_0^2}},
\nonumber
\end{eqnarray}
\noindent
where, according to Eq.~(\ref{eq10}),
\begin{equation}
D_0=\frac{2mc^2}{\hbar v_Fk_{\bot}}, \quad
B_0=\frac{\hbar v_Fk_{\bot}}{2k_BT}.
\label{eq13}
\end{equation}

The exact expressions (\ref{eq7}) and (\ref{eq9}) for the polarization
tensor at $l\geq 1$ are more complicated. Fortunately, much simpler
approximate expressions for them can be obtained in the
region of parameters
interesting from the experimental point of view. The {matter}  is that {for}
room temperature ($T=300\,$K) and at separations $a>100\,$nm already the first
Matsubara frequency satisfies the condition  $\xi_1\gg v_F/(2a)$.
Taking this {inequality} into account    and repeating the respective derivation of
Ref.~\cite{64} in our case of nonzero $m$ and $\mu$, one obtains
for $l\geq 1$
\begin{eqnarray}
&&
\Pi_{00,l}\approx\alpha\hbar \frac{ck_{\bot}^2}{\xi_l}\,
\left[\Psi\!\left(\frac{2mc^2}{\hbar\xi_l}\right)+
\tilde{Y}_l(T,m,\mu)\right],
\nonumber \\
&&
\Pi_{l}\approx\alpha\hbar \frac{\xi_lk_{\bot}^2}{c}\,
\left[\Psi\!\left(\frac{2mc^2}{\hbar\xi_l}\right)+
\tilde{Y}_l(T,m,\mu)\right],
\label{eq14}
\end{eqnarray}
\noindent
where
\begin{equation}
\tilde{Y}_l(T,m,\mu)=2\int_{2mc^2/(\hbar\xi_l)}^{\infty} du
\left(
\frac{1}{e^{B_lu+\frac{\mu}{k_BT}}+1}+
\frac{1}{e^{B_lu-\frac{\mu}{k_BT}}+1}\right)
\frac{ u^2+\left(\frac{2mc^2}{\hbar\xi_l}\right)^2}{u^2+1}.
\label{eq15}
\end{equation}
\noindent
We have performed
numerical computations of the Casimir force using the exact polarization
tensor (\ref{eq6}), (\ref{eq7}) and (\ref{eq9}) at all $l$ and,
alternatively, the exact expression (\ref{eq12}) at $l=0$ and the
approximate expressions (\ref{eq14}) at $l\geq 1$.
At $T=300\,$K, $a\geq 100\,$nm the obtained results
turned out to differ by less than 0.01\%.

Below we also consider the thermal correction to the Casimir force acting
between a Au sphere and a graphene sheet or graphene-coated substrate.
It is defined as
\begin{equation}
\Delta_T F(a,T)=F(a,T)-F(a,0).
\label{eq16}
\end{equation}
\noindent
The Casimir force at zero temperature, $F(a,0)$, is calculated by the
Lifshitz formula (\ref{eq1}) where summation in discrete Matsubara
frequencies is replaced with {an} integration over the imaginary frequency
axis according to
\begin{equation}
k_BT\sum_{l=0}^{\infty}{\vphantom{\sum}}^{\prime}\to
\frac{\hbar}{2\pi}\int_{0}^{\infty}d\xi.
\label{eq17}
\end{equation}
\noindent
{Along} with this {substitution}, the Matsubara frequencies $\xi_l$ in
Eqs.~(\ref{eq1})--(\ref{eq5}) are  replaced with $\xi$ and $q_l$,
$k_l^{(1)}$, $k_l^{(n)}$, $\Pi_{00,l}$, and $\Pi_l$
are {respectively} replaced with  $q$,
$k^{(1)}$, $k^{(n)}$, $\Pi_{00}$, and $\Pi$.

To calculate the reflection coefficients (\ref{eq3}) and (\ref{eq5})
at $T=0$ we need to find the limit\add{s} of  $\Pi_{00,l}^{(1)}$ and $\Pi_l^{(1)}$
{for} vanishing temperature. It is easily seen that the first fractions {among the round brackets},
{which contain} exponents in {the} denominators, on the right-hand sides of both
quantities in Eq.~(\ref{eq9}) {become zero} in the
limit $T\to 0$. As to the second fractions, {in the limit $T \rightarrow 0$} they
{become equal to unity  for
\begin{equation}
B_lu-\frac{\mu}{k_BT}<0{,}
\label{eq18}
\end{equation}
\noindent
{and vanish elsewhere.}
Taking into account that $u\geq D_l$, where $D_l$ is defined in Eq.~(\ref{eq10}),
{it follows} that in the limit $T\to 0$ the quantities $\Pi_{00}^{(1)}$ and
$\Pi^{(1)}$ are nonzero only {for} $mc^2<\mu$.
{Summing up the above considerations, in the limit $T \rightarrow 0$ one can replace the fractions between the round brackets in Eqs.~(\ref{eq9})  by  $\theta(\mu-mc^2)$, where $\theta(x)$ is the Heaviside step function equal to zero for $x\leq 0$
and unity for $x>0$, and restrict the integration over $u$ of the quantity between the square brackets to the interval $(2/\hbar)[mc/\tq,\mu/(c\tq)]$. After evaluating the latter elementary integral, and performing identical transformations, one arrives at the formula:}
\begin{eqnarray}
&&
\Pi_{00}^{(1)}(i\xi,k_{\bot},0,m,\mu)=\theta(\mu-mc^2)\left\{
\vphantom{\left[\left(\frac{\hbar\tq}{2mc}\right)\right]}
\frac{8\alpha c\mu}{v_F^2}-\frac{\alpha\hbar k_{\bot}^2}{\tq}
\left[
2M{\rm Im}(y_{m,\mu}\sqrt{1+y_{m,\mu}^2})+\frac{4mc}{\hbar\tq}
\right]\right.
\nonumber \\
&&~~\left.
-\frac{\alpha\hbar k_{\bot}^2}{\tq}(2-M)\left[
2{\rm Im}\ln(y_{m,\mu}+\sqrt{1+y_{m,\mu}^2})
-\pi+2\arctan\left(\frac{\hbar\tq}{2mc}\right)\right]\right\}.
\label{eq19}
\end{eqnarray}
\noindent
Here,  the following notations are introduced
\begin{equation}
M=1+\frac{4m^2c^2}{\hbar^2\tq^2}, \quad
y_{m,\mu}=\frac{\hbar\xi+2i\mu}{\hbar v_Fk_{\bot}\sqrt{M}}.
\label{eq20}
\end{equation}
\noindent
Then, for $\mu>mc^2$ {combining Eqs.~(\ref{eq6}), (\ref{eq7}), and (\ref{eq19})   one  obtains  the following expression for} the total 00-component of the polarization tensor  in the
limit $T\to 0$
\begin{eqnarray}
&&
\Pi_{00}(i\xi,k_{\bot},0,m,\mu)=
\frac{8\alpha c\mu}{v_F^2}-\frac{\alpha\hbar k_{\bot}^2}{\tq}
\left\{
2M{\rm Im}(y_{m,\mu}\sqrt{1+y_{m,\mu}^2})
\right.
\nonumber \\
&&~~\left.
+(2-M)\left[2{\rm Im}\ln(y_{m,\mu}+\sqrt{1+y_{m,\mu}^2})
-\pi\right]\right\}.
\label{eq21}
\end{eqnarray}
\noindent
Note that  {in the final expression (\ref{eq21})} two {of the} terms  in Eq.~(\ref{eq19}) exactly canceled against  the contribution of $\Pi_{00}^{(0)}$ {in} Eq.~(\ref{eq7}). In the alternative
case $mc^2\geq\mu$\add{,} the quantity $\Pi_{00}^{(1)}$ {in} Eq.~(\ref{eq19}) is {identically}
zero and the {complete expression of} $\Pi_{00}$ in the limit $T\to 0$ {reduces to}
\begin{equation}
\Pi_{00}(i\xi,k_{\bot},0,m,\mu)=\Pi_{00}^{(0)}(i\xi,k_{\bot},m),
\label{eq22}
\end{equation}
\noindent
i.e., {it} does not depend on $\mu$.

Similar results are obtained in the limit $T\to 0$ for the quantity $\Pi_l^{(1)}$
defined in Eq.~(\ref{eq9}). Calculating the integral over the same finite
interval, as for $\Pi_{00,l}^{(1)}$, after identical transformations one obtains
\begin{eqnarray}
&&
\Pi^{(1)}(i\xi,k_{\bot},0,m,\mu)=\theta(\mu-mc^2)\left\{
\vphantom{\left[\left(\frac{\hbar\tq}{2mc}\right)\right]}
-\frac{8\alpha \xi^2\mu}{cv_F^2}+2\alpha\hbar\tq k_{\bot}^2
\left[-\frac{2mc}{\hbar\tq}
+M{\rm Im}(y_{m,\mu}\sqrt{1+y_{m,\mu}^2})
\right.\right.
\nonumber \\
&&~~\left.\left.
-(2-M){\rm Im}\ln(y_{m,\mu}+\sqrt{1+y_{m,\mu}^2})
+(2-M)\arctan\left(\frac{2mc}{\hbar\tq}\right)\right]\right\}.
\label{eq23}
\end{eqnarray}
\noindent
In the case $\mu>mc^2$ the {complete expression of} $\Pi$ in the limit $T\to 0$ follows
from Eqs.~(\ref{eq6}), (\ref{eq7}) and (\ref{eq23})
\begin{eqnarray}
&&
\Pi(i\xi,k_{\bot},0,m,\mu)=
-\frac{8\alpha \xi^2\mu}{cv_F^2}+2{\alpha\hbar\tq k_{\bot}^2}
\left[\vphantom{\frac{\pi}{2}}
M{\rm Im}(y_{m,\mu}\sqrt{1+y_{m,\mu}^2})
\right.
\nonumber \\
&&~~\left.
-(2-M){\rm Im}\ln(y_{m,\mu}+\sqrt{1+y_{m,\mu}^2})
+\frac{\pi}{2}(2-M)\right].
\label{eq24}
\end{eqnarray}
\noindent
Here, again, the contribution of $\Pi^{(0)}$ {canceled against two  terms in the right-hand side of Eq.~(\ref{eq23})}.
In the alternative case $mc^2\geq\mu$ the contribution of $\Pi^{(1)}$
from Eq.~(\ref{eq23}) is equal to zero and one arrives at
\begin{equation}
\Pi(i\xi,k_{\bot},0,m,\mu)=\Pi^{(0)}(i\xi,k_{\bot},m),
\label{eq25}
\end{equation}

Thus, {we see from Eqs.~(\ref{eq22}) and (\ref{eq25}) that} in the limiting case of zero temperature {for} a sufficiently small
chemical potential satisfying the condition $\mu\leq mc^2$,
 the polarization tensor of graphene {in Eq.~}(\ref{eq7}) receives no corrections and therefore it depends
 only {on} the mass of quasiparticles. In order to
{affect} the polarization tensor of graphene in the limit $T\to 0$
(and, thus,  the reflection coefficients and  the Casimir force)
the chemical potential must satisfy the inequality $\mu>mc^2$.
In this case the additional terms to the polarization tensor of
graphene in the limit $T\to 0$ are given by Eqs.~(\ref{eq19}) and
(\ref{eq23}) and depend on both $\mu$ and $m$.
One can say that the case $\mu\leq mc^2$ corresponds to the interband transitions
when only $\Pi_{00}^{(0)}$ and $\Pi^{(0)}$ contribute to the polarization tensor.
For $\mu>mc^2$ the additional terms $\Pi_{00}^{(1)}$ and $\Pi^{(1)}$ in
the polarization tensor have to be considered, which means that the intraband
transitions come into play.

\section{The role of nonzero mass gap and chemical potential for
free-standing graphene}

In this section we calculate the Casimir force and {its}  thermal correction
for a Au sphere interacting with a graphene sheet characterized by nonzero
mass gap $m$ and chemical potential $\mu$. Calculations are performed at room
temperature $T=300\,$K over the separation region from 100\,nm to $1.5\,\mu$m
for {a} sphere radius $R=150\,\mu$m, as is typical for experiments
measuring the Casimir force \cite{73}.

To compute the Casimir force, one needs the value of $\ve_l^{(1)}$ for Au and
$m$ and $\mu$ for a graphene sheet. It {is well} known that  $\ve_l^{(1)}$
{can be determined on the basis of}  the tabulated optical data of Au \cite{74},
suitably extrapolated down to
zero frequency by either the lossy Drude or the lossless plasma model \cite{25,73}.
Although the Drude model, which takes into account the dissipation of free
electrons, may seem more realistic, all precise measurements of the Casimir
force between metallic test bodies turned out to be in  very good agreement with
the predictions of the Lifshitz theory using {the plasma model to}
{extrapolate} the optical data to
zero frequency  \cite{75,76,77,78,79,80,81,82}.
The {corresponding} predictions using the Drude model have been experimentally
excluded {nearly} at  100\% confidence level \cite{75,76,77,78,79,80,81,82}.

{A deep} theoretical understanding of why the lossless plasma model works
well at low frequencies in calculations of the fluctuation-induced Casimir force
is still missing. However, calculations of the Casimir force in graphene
systems, considered in this paper, are {unaffected by} the Drude-plasma dilemma which {only}
leads to  negligibly small differences in the obtained results \cite{53,67}.
The {reason is} {that the difference among the} theoretical predictions
{of} the Drude or plasma
models originates mainly from the TE zero-frequency contribution to the
Lifshitz formula (\ref{eq1}). {The latter contribution involves the {\it product} of the reflection coefficients $r_{\rm TE}^{(1)}(0,k_{\bot})$ and $R_{\rm TE}(0,k_{\bot})$ for Au and graphene.}   If the Drude model is used, the reflection
coefficient $r_{\rm TE}^{(1)}(0,k_{\bot})$ {of Au} is found to be zero,
{while a nonzero result is obtained}  if the
plasma model is employed. {Since, however, the reflection coefficient
$R_{\rm TE}(0,k_{\bot})$ for graphene is negligibly small}
 due to the smallness of $\Pi_0$ in
Eq.~(\ref{eq12}),{ it follows at once that the value of the reflection coefficient
$r_{\rm TE}^{(1)}(0,k_{\bot})$  is irrelevant, whether it is zero or nonzero. As a result, the Drude and the plasma models lead to practically undistinguishable values for} the Casimir force
{acting} between {a} metallic test body and graphene. Below an experimentally consistent
extrapolation of the optical data for Au to zero frequency by means of the
plasma model \cite{25,73} is used in all computations.

Now we discuss possible values of the mass gap $mc^2$ and the chemical potential
$\mu$. It is common knowledge that in  pristine graphene the Dirac-type
electronic excitations are massless. {However, definite}  conditions {existing in real samples}, such as
electron-electron interactions,  structure {defects}, the presence of a substrate,
and some other effects give rise to a nonzero mass gap \cite{36,83,84,85}.
The exact value of $mc^2$ for a specific graphene sample usually remains unknown.
Realistic {estimates bound} the mass gap {to the region} $mc^2<0.1\,$eV for
free-standing graphene, while  for a graphene sheet deposited on a substrate  the bound is $mc^2<0.2\,$eV.

Similar to the mass-gap parameter, for  pristine graphene the chemical
potential is equal to zero. For real graphene samples, however, there is always
some fraction of extraneous atoms, i.e., real graphene samples are always doped
with some doping concentration $n$. At zero temperature the respective chemical
potential is given by \cite{86}
\begin{equation}
\mu(T=0)=\hbar v_F\sqrt{\pi n}.
\label{eq26}
\end{equation}
In so doing it is almost independent on the temperature \cite{86}.
For nearly undoped graphene films in high vacuum used in the {Casimir} experiment {\cite{35}}
the value of $n\approx 1.2\times 10^{10}\,\mbox{cm}^{-2}$ was estimated {based} on
measurements of two-dimensional mobility.
The {corresponding} maximum
value of the chemical potential obtained from Eq.~(\ref{eq26})
is $\mu=0.02\,$eV.
As two more examples, the values of chemical potential for doping concentrations
$n\approx 7.5\times 10^{11}$ and $2\times 10^{13}\,\mbox{cm}^{-2}$ are equal to
$\mu=0.1$ and 0.5\,eV, respectively. Note also that in {the} measurement of the optical
conductivity of graphene {reported in} Ref.~\cite{86a} a representative value of $\mu=0.1\,$eV was used for
a graphene sheet on the top of a SiO${}_2$ substrate.

Computations of the Casimir force between a Au sphere and a free-standing graphene
sheet {are conveniently done} using the dimensionless variables
\begin{equation}
y=2aq_l,\quad \zeta_l=\frac{2a\xi_l}{c}.
\label{eq27}
\end{equation}
\noindent
In terms of these variables the Lifshitz formula (\ref{eq1})
takes the form
\begin{equation}
F(a,T)=\frac{k_BTR}{4a^2}
\sum_{l=0}^{\infty}{\vphantom{\sum}}^{\prime}
\int_{\zeta_l}^{\infty}\!\!ydy
\sum_{\alpha}
\ln\left[1-r_{\alpha}^{(1)}(i\zeta_l,y)R_{\alpha}^{(n)}(i\zeta_l,y)
e^{-y}\right].
\label{eq28}
\end{equation}
\noindent
Here, the reflection coefficients (\ref{eq2}) {for} a Au surface take the form
\begin{eqnarray}
&&
r_{\rm TM}^{(1)}(i\zeta_l,y)=\frac{\varepsilon_l^{(1)}y-
\sqrt{y^2+(\ve_l^{(1)}-1)\zeta_l^2}}{\varepsilon_l^{(1)}y+
\sqrt{y^2+(\ve_l^{(1)}-1)\zeta_l^2}},
\nonumber \\
&&
r_{\rm TE}^{(1)}(i\zeta_l,y)=
\frac{y-\sqrt{y^2+(\ve_l^{(1)}-1)\zeta_l^2}}{y+\sqrt{y^2+
(\ve_l^{(1)}-1)\zeta_l^2}}.
\label{eq29}
\end{eqnarray}
\noindent
The reflection coefficients (\ref{eq5}) {for} a graphene sheet are given by
\begin{eqnarray}
&&
R_{\rm TM}(i\zeta_l,y)=
\frac{y\tilde{\Pi}_{00,l}}{y\tilde{\Pi}_{00,l}+2(y^2-\zeta_l^2)},
\nonumber \\
&&
R_{\rm TE}(i\zeta_l,y)=-
\frac{\tilde{\Pi}_{l}}{\tilde{\Pi}_{l}+2y(y^2-\zeta_l^2)},
\label{eq30}
\end{eqnarray}
\noindent
where the dimensionless polarization tensor is defined as
\begin{equation}
\tilde{\Pi}_{00,l}=\frac{2a{\Pi}_{00,l}}{\hbar}, \quad
\tilde{\Pi}_{l}=\frac{(2a)^3{\Pi}_{l}}{\hbar}.
\label{eq31}
\end{equation}
\noindent
In doing so the quantities (\ref{eq12}), (\ref{eq14}), (\ref{eq21}), and
(\ref{eq24}) are also rewritten in terms of the dimensionless variables (\ref{eq27}).

First, we investigate the relative impact of nonzero $m$ and $\mu$ on the
Casimir force, as compared to the case of  pristine graphene
with $mc^2=\mu=0$. For this purpose we calculate
the quantity
\begin{equation}
\delta_{m,\mu} F=\frac{F(a,\,T,\,m,\,\mu)-
F(a,\,T,\,0,\,0)}{F(a,\,T,\,0,\,0)}.
\label{eq32}
\end{equation}
\noindent
Numerical computations have been performed at $T=300\,$K by using
Eqs.~(\ref{eq28})--(\ref{eq30}), (\ref{eq12}) and (\ref{eq14}).
The computational results in percents as functions of separation
are shown in Fig.~\ref{fg1}(a) by
the {five} lines 1, 2, 3, 4, and 5 {corresponding, respectively, to the following five combinations of values of $m$ and $\mu$:}
$m=0,\>\mu=0.5\,$eV; $m=0,\>\mu=0.1\,$eV;
$mc^2=0.1\,\mbox{eV},\>\mu=0$;
$mc^2=0.15\,\mbox{eV},\>\mu=0$;
and $mc^2=0.2\,\mbox{eV},\>\mu=0$.
[Note that Figs.~\ref{fg1}(b) and \ref{fg1}(c), {which refer} to the case
of graphene deposited on a substrate, are discussed in Sec.~IV.]
As is seen in Fig.~\ref{fg1}(a), the presence of nonzero mass gap and
chemical potential acts on the Casimir force in opposite directions.
The magnitude of the Casimir force decreases with increasing $m$, { while it}
increases with increasing $\mu$.
This result finds \add{a} simple physical explanation.
An increase of the chemical potential
essentially increases graphene's conductivity, so that one should expect
the force to grow. On the other hand, by increasing the mass gap one is lowering
the mobility, which in turn lowers the conductivity and brings the force down.
The relative impact of both parameters
decreases with increasing separation distance between the sphere and
the graphene sheet.

We next consider the dependence of the Casimir force between a Au sphere
and a graphene sheet and {its} thermal correction  (\ref{eq16}) on the
value of {the} mass gap {for} different values of the chemical potential.
{When doing that,} the Casimir force $F(a,0)$ is computed using the Lifshitz
formula at zero temperature, i.e. Eq.~(\ref{eq1}) {with the replacement} (\ref{eq17}), together with the {expressions for the}
polarization tensor in Eqs.~(\ref{eq21}), (\ref{eq22}), (\ref{eq24}), and
(\ref{eq25}). The computational results for the Casimir force $F$ are
shown {for} $a=0.1\,\mu$m {and} $T=300\,$K in Fig.~\ref{fg2}(a)
 as functions of the mass gap for {three different} values of chemical potential
$\mu=0$, 0.1, and 0.5\,eV, {corresponding to lines 1, 2 and 3}, respectively.  {The computational results for the thermal correction to
the Casimir force, $\Delta_T F$, and for the fractional weight of the thermal correction
in the total force, $\Delta_T F/F$, are displayed in Figs.~\ref{fg2}(b) and
\ref{fg2}(c) for the same values of the parameters as in Fig.~\ref{fg2}(a)}.

As is seen in Fig.~\ref{fg2}(a), the magnitude of the Casimir force
decreases monotonously with increasing $m$ and increases with an increase
of $\mu$ in accordance with Fig.~\ref{fg1}(a). The values of $F$ in
Fig.~\ref{fg2}(a) at $m=0$ are in agreement with the computational results of
Ref.~\cite{66}, where the enhanced thermal Casimir force {for} nonzero $\mu$
was found. The dependence of the thermal correction to the Casimir force
on $m$ in Figs.~\ref{fg2}(b) and \ref{fg2}(c) is nonmonotonous (this
effect was already
noted in Ref.~\cite{53} for the case $\mu=0$).
The characteristic {discontinuity displayed  by} the derivative
{of the $\mu=0.1\,$eV curves in Figs.~\ref{fg2}(b) and \ref{fg2}(c)  } { for} $mc^2=0.1\,$eV
{is} explained by the fact that for
$\mu=0.1\,$eV {the} contributions $\Pi_{00}^{(1)}$ and $\Pi^{(1)}$ to the polarization
tensor at $T=0$ {both turn into zero}  for  $mc^2\geq 0.1\,$eV (see Sec.~II).
Note that for  free-standing graphene {values of} $mc^2$ exceeding 0.1\,eV
{are} somewhat {unrealistic}.
This region is shown {in Figs.~\ref{fg2}(b) and \ref{fg2}(c) only} for comparison {purposes} with the case of
graphene deposited on a substrate (see Sec.~IV). On the whole, under the
condition $mc^2<\mu$ the thermal effect decreases in magnitude with increasing
chemical potential, and its relative role in the Casimir force drops down.

For comparison purposes, the {$300\,$K} thermal Casimir force between a Au sphere and
a free-standing graphene sheet [Fig.~\ref{fg3}(a)], {its} thermal correction
[Fig.~\ref{fg3}(b)], and the fractional weight of the thermal correction
in the total thermal force [Fig.~\ref{fg3}(c)]
are shown {as functions of the mass gap} {for the larger} separation $a=1\,\mu$m
 for {three} different
values of the chemical potential
$\mu=0$, 0.1, and 0.5\,eV {corresponding}, respectively, {to lines 1, 2 and 3 in the figures}.
As {it can be} seen {by a} comparison of   Figs.~\ref{fg2} and \ref{fg3},
 both the Casimir force and {its} thermal correction {have a similar qualitative behavior}
{for the two} different separations. Although the magnitudes of
the thermal correction at $a=1\,\mu$m are smaller than {those found for} $a=0.1\,\mu$m
[compare Figs.~\ref{fg3}(b) and \ref{fg2}(b)],
{they represent larger} fractions {of} the total
force at $a=1\,\mu$m [see Figs.~\ref{fg3}(c) and \ref{fg2}(c)].
As is seen in Fig.~\ref{fg3}(c), for a sufficiently large  mass gap
{a} chemical potential not exceeding 0.1\,eV makes almost no impact on the
fractional weight of the thermal correction in the total Casimir force.

\section{Suppressed impact of the  mass gap and chemical {\protect \\} potential for
graphene-coated dielectric substrates}

Here, we consider the Casimir force and its thermal correction for a Au-coated
sphere of $R=150\,\mu$m radius interacting with a real graphene sheet deposited on
a dielectric plate. Numerical computations were performed at $T=300\,$K by using
the Lifshitz formula (\ref{eq28}) for plates made of SiO${}_2$ (vitreous silica)
and high-resistivity Si. The reflection coefficients (\ref{eq3}) are expressed
in terms of dimensionless variables (\ref{eq27}) as
\begin{eqnarray}
&&
R_{\rm TM}^{(n)}(i\zeta_l,y)=\frac{\varepsilon_l^{(n)}y(y^2-\zeta_l^2)+
\sqrt{y^2+(\ve_l^{(n)}-1)\zeta_l^2}\left[y\tilde{\Pi}_{00,l}-
(y^2-\zeta_l^2)\right]}{\varepsilon_l^{(n)}y(y^2-\zeta_l^2)+
\sqrt{y^2+(\ve_l^{(n)}-1)\zeta_l^2}\left[y\tilde{\Pi}_{00,l}+
(y^2-\zeta_l^2)\right]},
\nonumber \\
&&
R_{\rm TE}^{(n)}(i\zeta_l,y)=
\frac{(y^2-\zeta_l^2)[y-\sqrt{y^2+(\ve_l^{(n)}-1)\zeta_l^2}]-
\tilde{\Pi}_l}{(y^2-\zeta_l^2)[y+\sqrt{y^2+(\ve_l^{(n)}-1)\zeta_l^2}]+
\tilde{\Pi}_l}.
\label{eq33}
\end{eqnarray}
\noindent
We begin with the case of a SiO${}_2$ substrate.

\subsection{Silica plate}

The dielectric permittivity of SiO${}_2$ along the imaginary frequency axis,
$\ve_l^{(2)}\equiv\ve^{(2)}(i\xi_l)$, {can be described very accurately by a simple analytic formula}
\cite{25,87,87a}.  The four numerical coefficients    involved in this formula
have been determined from a fit to the tabulated optical data of SiO${}_2$
\cite{74}. The resulting static dielectric permittivity of SiO${}_2$ is
$\ve_0^{(2)}=3.81$.
Numerical computations of the Casimir force, $F$, and its thermal correction,
$\Delta_T F$, have been performed at $T=300\,$K  using Eqs.~(\ref{eq28}),
(\ref{eq29}), (\ref{eq33}) {together with the} expressions for the polarization
tensor of graphene presented in Sec.~II for both nonzero and zero temperature.

First, we calculate the relative {difference $\delta_{m,\mu} F$ among}   the Casimir forces between
a Au sphere and a SiO${}_2$ plate coated either with real graphene characterized
by nonzero $m$ and $\mu$\add{,} or with  pristine graphene for which
$mc^2=\mu=0$ [see Eq.~(\ref{eq32})]. The computational results for $\delta_{m,\mu}F$
in percents are presented in Fig.~\ref{fg1}(b) as  functions of separation
by the five lines 1, 2, 3, 4, and 5 {corresponding to} the following combinations of parameters:
$m=0,\>\mu=0.5\,$eV; $m=0,\>\mu=0.1\,$eV;
$mc^2=0.1\,\mbox{eV},\>\mu=0$;
$mc^2=0.15\,\mbox{eV},\>\mu=0$;
and $mc^2=0.2\,\mbox{eV},\>\mu=0$, respectively.
As is seen in Fig.~\ref{fg1}(b), for the graphene-coated substrate the  presence
of {a} nonzero mass gap and chemical potential {has an opposite effect on} the magnitude of the Casimir
force {similar to the case} of a free-standing graphene
sheet [see Fig.~\ref{fg1}(a)]. The striking difference between the two cases is,
however, that for a graphene-coated substrate the impact of nonzero $m$ and $\mu$
is up to an order of magnitude smaller than for a free-standing graphene
sheet.
This is explained by the fact that the substrate material {provides} the dominant
contribution to the reflection coefficient of the graphene-coated substrate
(see a discussion in the next paragraph for more details).
It is notable also that for the largest mass
considered ($mc^2=0.2\,$eV) the magnitude  of $\delta_{m,\mu}F$ increases with
increasing separation, while this phenomenon is not observed for  free-standing graphene
with any mass.

Next, we present the computational results for the Casimir force between
a Au sphere and a
graphene-coated SiO${}_2$ plate {for} $T=300\,$K, $a=0.1\,\mu$m [Fig.~\ref{fg4}(a)],
its thermal correction [Fig.~\ref{fg4}(b)], and the fractional weight of the
thermal correction in the Casimir force [Fig.~\ref{fg4}(c)] as functions of the
mass gap for $\mu=0$, 0.1, and 0.5\,eV (lines 1, 2, and 3, respectively).
As is seen from the comparison of Figs.~\ref{fg2}(a)--\ref{fg2}(c) and
\ref{fg4}(a)--\ref{fg4}(c),
the Casimir force, its thermal correction, and the fractional weight of the
thermal correction in the  force {have the same qualitative behavior in}
the cases of a free-standing
graphene and for a graphene-coated SiO${}_2$ substrate. One should note, however,
that for a graphene-coated SiO${}_2$ plate the magnitude of the Casimir force is
several times larger than for a free-standing graphene [see  Figs.~\ref{fg2}(a)
and \ref{fg4}(a)].
Physically this is explained by the fact that for a graphene-coated substrate
the reflection coefficient is much larger than for a free-standing graphene sheet.
Specifically, for a free-standing graphene the reflection coefficient is
proportional to the fine structure constant. Upon including a substrate, the
overall reflection coefficient of the substrate and graphene system is dominated
by the large contribution from the substrate material. The latter depends on its
dielectric permittivity and corresponds to the zeroth order term in an
expansion in powers of the fine structure constant.
The magnitude of the thermal correction for a graphene-coated
substrate is somewhat smaller and its fractional weight in the total Casimir
force is much smaller than for a free-standing graphene sheet. This is seen from
the comparison of Fig.~\ref{fg2}(b) with Fig.~\ref{fg4}(b) and Fig.~\ref{fg2}(c)
with Fig.~\ref{fg4}(c), respectively. One can say that a substrate acts on the
zero-temperature force and on the thermal correction to it in opposite
directions.

Similar to Fig.~\ref{fg4}, in Fig.~\ref{fg5} the computational results for $F$,
$\Delta_TF$ and $\Delta_TF/F$ in the case of a graphene-coated SiO${}_2$ plate
are presented at $T=300\,$K, but at the larger sphere-plate separation
$a=1\,\mu$m. All  notations in Fig.~\ref{fg5} are the same as in Fig.~\ref{fg4}.
As is seen in Fig.~\ref{fg5}, the {qualitative features of the} dependence of
all considered
quantities on the mass-gap parameter {do not change}.
The magnitude of the Casimir force in Fig.~\ref{fg5}(a) is larger and the
magnitude of its thermal correction in Fig.~\ref{fg5}(b) is smaller than
in Figs.~\ref{fg3}(a) and \ref{fg3}(b) plotted at the same separation for
a free-standing graphene sheet. The fractional weight of the thermal correction
in the total force in Fig.~\ref{fg5}(c)  is up to a factor of 3 smaller than
in Fig.~\ref{fg3}(c). These results remain valid for any considered value of
the chemical potential and mass-gap parameter.

\subsection{Silicon plate}

The dielectric permittivity of high-resistivity Si along the imaginary frequency
axis $\ve_l^{(3)}\equiv \ve^{(3)}(i\xi_l)$ was obtained from the optical data
for its complex index of refraction \cite{74}. Unlike SiO${}_2$, Si possesses
{a} rather large static dielectric permittivity $\ve_0^{(3)}=11.66$,
and this influences both the Casimir force and its thermal correction.
All computations are performed using the same equations as in Sec.~IVA.

Specifically, the relative deviations of the Casimir force between
a Au sphere and a Si plate coated with either real or pristine graphene
as functions of separation are shown in Fig.~\ref{fg1}(c) by the five lines
for the same pairs  ($mc^2,\,\mu$) for real graphene as in Figs.~\ref{fg1}(a)
and \ref{fg1}(b). It is again seen that nonzero $m$ and $\mu$ influence the
Casimir force in  opposite directions. This impact, however, is further
decreased as compared to the case of SiO${}_2$ plate.

In Figs.~\ref{fg6}(a)--\ref{fg6}(c) the computational results for the Casimir
force between
a Au sphere and a graphene-coated Si plate, for its thermal correction and
for the fractional weight of the thermal correction in the force are
presented for $a=0.1\,\mu$m, $T=300\,$K as functions of the mass gap by the
lines 1, 2, and 3 plotted for $\mu=0$, 0.1, and 0.5\,eV, respectively.
As is seen in Fig.~\ref{fg6}(a), the magnitudes of the Casimir force are
much larger than for the case of {the} SiO${}_2$ plate shown in Fig.~\ref{fg4}(a),
 as a result of  the larger {value} $\ve_0^{(3)}$ {of the static permittivity} of Si as compared to  $\ve_0^{(2)}$ of SiO${}_2$.
{}From Figs.~\ref{fg6}(b) and \ref{fg6}(c) it follows, however, that both
the thermal correction and its fractional weight in the total force are
smaller than for a SiO${}_2$ plate. Hence, for substrates with larger $\ve_0$
the impact of graphene coating with nonzero $m$ and $\mu$ becomes smaller.
This effect was {reported}
in Ref.~\cite{60} {for dielectric plates coated with pristine graphene}.

{The analogous}  computational results for $F$,
$\Delta_TF$ and $\Delta_TF/F$ {for} the larger  separation {$a=1\,\mu$m} between
a Au sphere and a graphene-coated Si plate
 are shown in Figs.~\ref{fg7}(a)--\ref{fg7}(c).
Notations {are the same }as in Fig.~\ref{fg6}.
{}From Fig.~\ref{fg7}(a) one can see that the
magnitudes of the Casimir force are much larger than those {shown}
in Fig.~\ref{fg5}(a),  which refer to
the case of a graphene-coated SiO${}_2$ plate.
The magnitudes of the thermal correction in Fig.~\ref{fg7}(b) and
 its fractional weight in the total force in Fig.~\ref{fg7}(c)  are
{smaller},  compared to the case of {a} SiO${}_2$ substrate at the same
 separation distance [see Figs.~\ref{fg5}(b) and \ref{fg5}(c)].


\section{Role of nonzero mass gap and chemical potential in differential
measurements of the thermal effect for graphene}

It has long been known that differential measurements allow to achieve
a very high precision and reliability. In Casimir physics they were used
to measure the optically modulated Casimir force between a Au sphere and
a Si plate illuminated with laser pulses \cite{88} and in the so-called
Casimir-less {experiments searching} for  Yukawa-type corrections to Newton's
gravitational law \cite{89,90}. Recently it was found that  differential
measurement schemes allow {for} a {huge amplification} of the difference
{among the} theoretical
predictions of the Lifshitz theory using either the Drude or the plasma
model {in the} extrapolation of  optical data down to zero frequency
\cite{91,92,93}.  Using one of these {differential} schemes, {in which} the difference
{among the} alternative theoretical predictions
{could be made} as large as {by} a factor of 1000, {the experiment of Ref.~\cite{82} confirmed} the plasma model approach to the
Casimir force between metallic test bodies  and {undeniably excluded the} Drude
model.
This result {provided further support to} several previous measurements in favor of the plasma
model \cite{75,76,77,78,79,80,81}, {in which, however,} the predicted difference between the
two approaches did not exceed a few percent. {Very} recently, {a} universal
differential measurement  {has been} proposed which allows {for a} clear discrimination
between different theoretical approaches to the Casimir force not only for
metallic, but {also} for dielectric test bodies \cite{94}.

In this section, we consider the role of  nonzero mass gap and  chemical
potential in the  differential {setup aiming at the observation} of the giant thermal
effect in graphene systems,  proposed by us in Ref.~\cite{67}. {In that work it} was shown that the thermal effect
can be observed, using {a feasible adaptation of a currently} available experimental setup,
by measuring
the difference among the Casimir forces between a Au-coated sphere and
the two halves of a Si plate, one of which is coated and the other is
uncoated with a graphene sheet \cite{67}. {Since} in
Ref.~\cite{67} the parameters  {for} pristine graphene have been used
in {the} computations, it is now important to check whether } the {encouraging}
conclusions {reached there} concerning the feasibility of this experiment remain valid
for real graphene {samples which posses} nonzero mass gap and chemical potential.

{We recall that in Ref.~\cite{67} we proposed to measure the differential force}
\begin{equation}
F_{\rm diff}(a,T)=F_{\rm Si}(a,T)-F(a,T),
\label{eq34}
\end{equation}
\noindent
where $F(a,T)$ is the Casimir force between a Au-coated sphere and
a graphene-coated Si plate expressed by Eqs.~(\ref{eq28}), (\ref{eq29}),
and (\ref{eq33}), and $F_{\rm Si}(a,T)$ is the Casimir force between
a Au sphere and uncoated Si plate. The latter {force} is also expressed by the
Lifshitz formula (\ref{eq28}), in which the reflection coefficients (\ref{eq29})
{remain unchanged}.
The reflection coefficients $R_{\alpha}^{(3)}$ {are} replaced with
 coefficients of the uncovered Si plate  having a form analogous to Eq.~(\ref{eq29})
 in which, however, the dielectric permittivity
$\ve_l^{(1)}$ of Au is {replaced by} the permittivity
$\ve_l^{(3)}$ of Si.

\begin{figure}[b]
\vspace*{-1cm}
\centerline{\hspace*{2.5cm}
\includegraphics{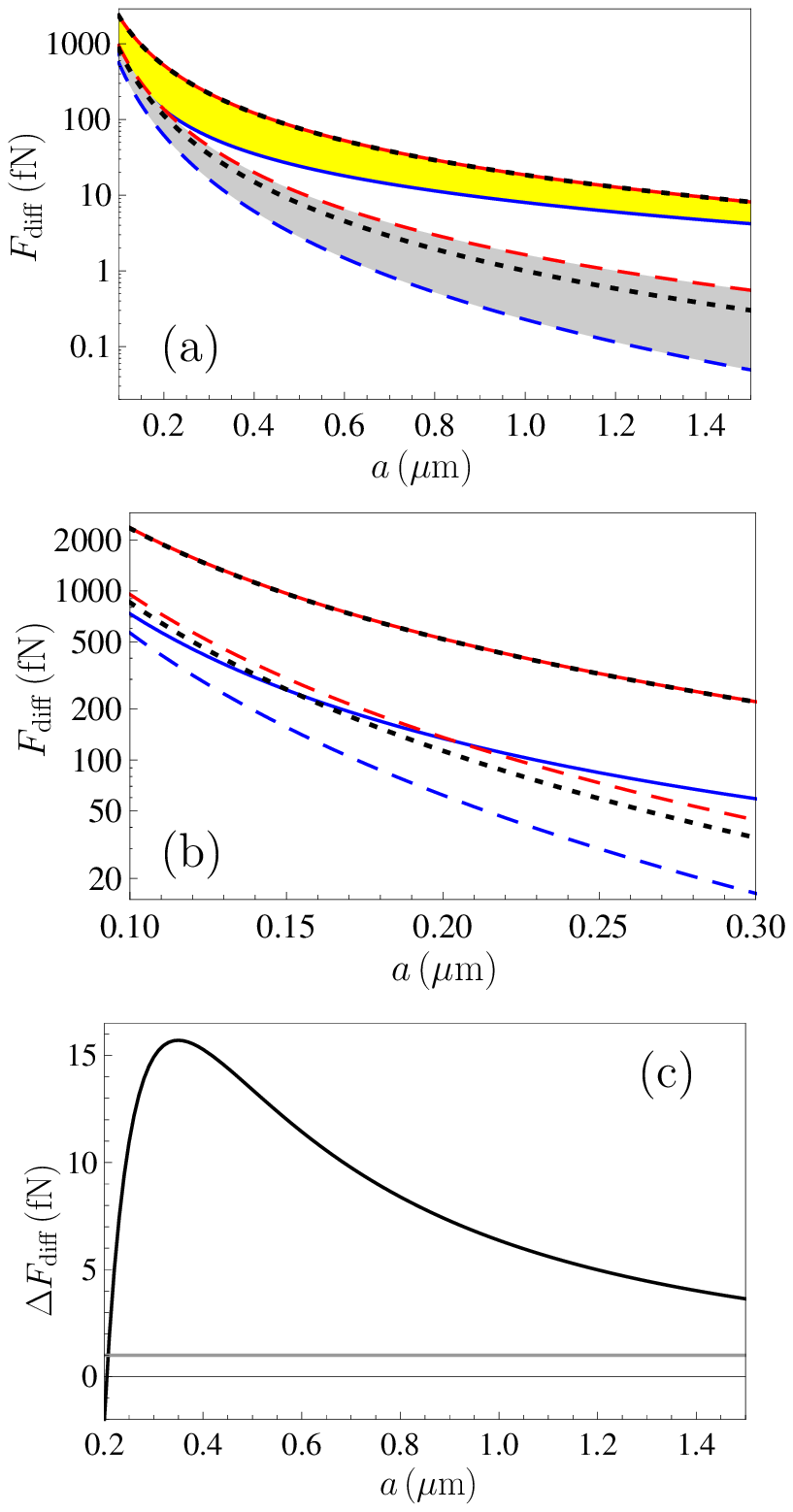}
}
\vspace*{-12cm}
\caption{\label{fg8}
(a) The difference {among} the Casimir forces between a Au sphere
and uncoated and graphene-coated Si plate
is shown as a function of
separation. The top and bottom solid lines are plotted at $T=300\,$K
for $mc^2=0,\,\mu=0.02\,$eV and for
$mc^2=0.2\,\mbox{eV},\,\mu=0$, respectively. The top and
bottom dashed lines are plotted at $T=0\,$K for the same respective pairs
of {values for} $mc^2$ and $\mu$. The top and bottom dotted lines are
plotted for the pristine graphene at $T=300\,$K and $T=0\,$K,
respectively.
(b) The region of short separations  from (a) is shown on an
enlarged scale.
(c) The difference between the bottom solid and top dashed lines
in (a)  is shown  by the top solid line in comparison with the
experimental error (the horizontal line) as a function of separation.
}
\end{figure}

Using these formulas we have performed numerical computations of $F_{\rm diff}$
as a function of separation. The computational results for $T=300\,$K are shown
in Fig.~\ref{fg8}(a) by the top and bottom solid lines plotted for
$mc^2=0$, $\mu =0.02\,$eV and $mc^2=0.2\,$eV, $\mu =0$, respectively.
The top and bottom dashed lines in this figure {show} the computational
results for  $F_{\rm diff}$ at $T=0$ for the same {two combinations}
 of { values}
($m,\,\mu$). We remind that  $\mu=0.02\,$eV
{represents the value of the chemical potential}  for {the} graphene
samples used in {the} experiment \cite{35}, {which measured} the gradient of the Casimir
force between a Au-coated sphere and graphene-coated substrate, whereas
$mc^2=0.2\,$ {represents} the  estimated {upper bound on the} value of $mc^2$ for graphene-coated
substrates. {Therefore,} one can {confidently} conclude  that {for realistic samples} the thermal
correction to the differential force (\ref{eq34}),
\begin{equation}
\Delta_T F_{\rm diff}(a,T)=F_{\rm diff}(a,T)-F_{\rm diff}(a,0),
\label{eq35}
\end{equation}
\noindent
belongs to the band defined by the difference between the two bands bounded,
respectively, by the two pairs of solid and
dashed lines in Fig.~\ref{fg8}(a). The upper bound on the band for $\Delta_T F_{\rm diff}$
is therefore equal to the difference  among the top solid and bottom dashed lines
in Fig.~\ref{fg8}(a),
while its lower bound is equal to the difference among the bottom solid and
top dashed lines of that figure.

The top and bottom dotted lines in Fig.~\ref{fg8}(a) {were} computed
{supposing that} half of {the} Si plate is coated not with {a} real, but {rather} with a pristine
graphene for $T=300\,$K and $T=0$, respectively. The difference {among} these lines
was used as an estimate of $\Delta_T F_{\rm diff}$ in Ref.~\cite{67}.
As is seen in  Fig.~\ref{fg8}(a), the top solid line ($T=300\,$K, $mc^2=0$,
$\mu =0.02\,$eV) and the top dotted line ($T=300\,$K, pristine graphene)
are almost coinciding. This {indicates} that the chemical potential $\mu=0.02\,$eV
does not influence the value of the Casimir force. For {a} better visualization,
in Fig.~\ref{fg8}(b) the region of short separations is shown on an enlarged
scale.

Now we are in a position to determine the feasibility of the proposed
experiment on measuring the giant thermal effect in graphene systems with
{full} account of {the} nonzero mass gap and chemical potential {of real samples}.
For this purpose,
in Fig.~\ref{fg8}(c) we plot the minimum value of the thermal correction in
$F_{\rm diff}$,
\begin{equation}
\Delta F_{\rm diff}(a,T)=\min\Delta_T F_{\rm diff}(a,T),
\label{eq36}
\end{equation}
\noindent
{which is} equal to the difference between the bottom solid and top dashed lines
in Fig.~\ref{fg8}(a). The horizontal line in Fig.~\ref{fg8}(c) indicates
the {magnitude} of the experimental error of 1\,fN  {achieved in a}
differential measurement of the Casimir force
{that has been performed already} \cite{82}.
As is seen in Fig.~\ref{fg8}(c), {even taking into  account the mass gap and
chemical potential,} the minimum possible value of the thermal
correction $\Delta F_{\rm diff}$  far exceeds the experimental error over the wide
separation region from 220\,nm to $1.5\,\mu$m. Thus, the proposed
experiment allows {for} a reliable observation of the giant thermal effect
for real graphene samples.

\section{Conclusions and discussion}

In the foregoing we have investigated the impact of nonzero mass gap and
chemical potential on the Casimir force and its thermal correction for
graphene systems. The experimentally relevant configurations of a Au-coated
sphere interacting with either a free-standing graphene sheet or
graphene-coated dielectric substrates made of different materials have been the
focus of our attention. It was found that the  mass gap and
chemical potential produce pronounced effects on both the Casimir force and
its thermal correction, indicating that both quantities should be taken into account in  investigations of
 fluctuation-induced phenomena in graphene and other 2D-materials, as well as
  in
{prospective} applications of these phenomena {to} nano- and micromechanical systems.

According to our results, {which were derived}  in the framework of {the} rigorous QED approach
{based on} the exact  {finite-temperature} polarization tensor of graphene,
the mass gap and chemical potential act {in opposite directions} on the magnitude of the
Casimir force. Specifically, it was shown that with
increasing $m$ and fixed $\mu$ the magnitude of the Casimir force decreases,
whereas it increases with increasing $\mu$ and fixed $m$. This {behavior, which} is
{observed}
both {for a} free-standing {graphene sheet and  for graphene}  deposited on a substrate, leads to {a} partial compensation of both effects when the Casimir force is
{worked out taking into account the simultaneous}
 influence of nonzero $m$ and $\mu$.
 Although, as discussed in Secs.~III and IV, the above results are quite expected on
 physical grounds, {a} precise quantitative evaluation
 {requires the use of the} formalism {presented  in this work}.

 Another important property demonstrated in this paper is that the impacts of
  nonzero mass gap and chemical potential on the Casimir force for
 graphene-coated plates are {both} much smaller than for a free-standing graphene
 sheet. Since most applications {use}  graphene-coated
 substrates, {our results suggest that} the
 {possibility of} controlling the Casimir force by means of the
 chemical potential {is} somewhat problematic.

Our computational results show that {the magnitude of the Casimir force} for graphene-coated substrates {is much larger than for}
a free-standing graphene, but the thermal correction and its fractional weight
in the total force {are smaller}. The computations made for SiO${}_2$ and
Si plates demonstrate that the influence of the graphene coating on the Casimir force
and its thermal correction decreases when the substrate has a higher
static dielectric permittivity.
Our investigation of the { dependence of the} thermal correction {on the} mass gap
for different values of {the} chemical potential revealed
{the existence of a discontinuity of the} derivative {of the Casimir force with respect to $mc^2$}
{for} $mc^2=\mu$.
{The origin of this discontinuity resides in}  the fact that
{for values of the chemical potential such that} $\mu\leq mc^2$
{the zero-temperature polarization tensor of graphene is independent of} $\mu$.

Finally, we have determined the role of nonzero $m$ and $\mu$ in
{a differential} experiment {recently proposed by us} \cite{67} {to observe} the giant
thermal effect in the Casimir force at short separations. For this purpose,
the differential Casimir force between a Au sphere and {the} two halves of a Si
plate, one uncoated and the {other} coated with graphene, was computed as a function
of separation, taking  into account {the influence of}
 $m$ and $\mu$. {We estimated} the minimum value of the thermal
correction {to} the Casimir force  as a function of separation
 with the same value of the chemical potential as for graphene samples in a recent experiment probing the Casimir force between a Au sphere and a graphene-coated
  substrate \cite{35}. In this computation we conservatively used the maximum possible
value  for  the mass-gap parameter.
Comparison with the
{force sensitivity of state-of-the-art differential Casimir setups}
 {confirms} the feasibility of the proposed experiment with  account {of} $m$ and $\mu$.

The  formalism for {the} calculation of the thermal Casimir force in
graphene systems accounting for nonzero $m$ and $\mu$, presented in this work, opens {many}
novel
opportunities in both fundamental and applied research.
{For example,} it {may} allow {for a} determination of the mass gap of
{a} graphene sheet
by fitting the experimental data for {the} giant thermal effect to the
{theoretical} results obtained with different values of $m$.
A {direct} observation of the giant thermal effect in graphene, combined
with first-principle computations,  might open novel opportunities
for {the} modification and control of the Casimir force in graphene-based
micromechanical systems.

\section*{Acknowledgments}
The work of V.M.M. was partially supported by the Russian
Government
Program of Competitive Growth of Kazan Federal University.

\newpage
\begin{figure}[b]
\vspace*{-1cm}
\centerline{\hspace*{2.5cm}
\includegraphics{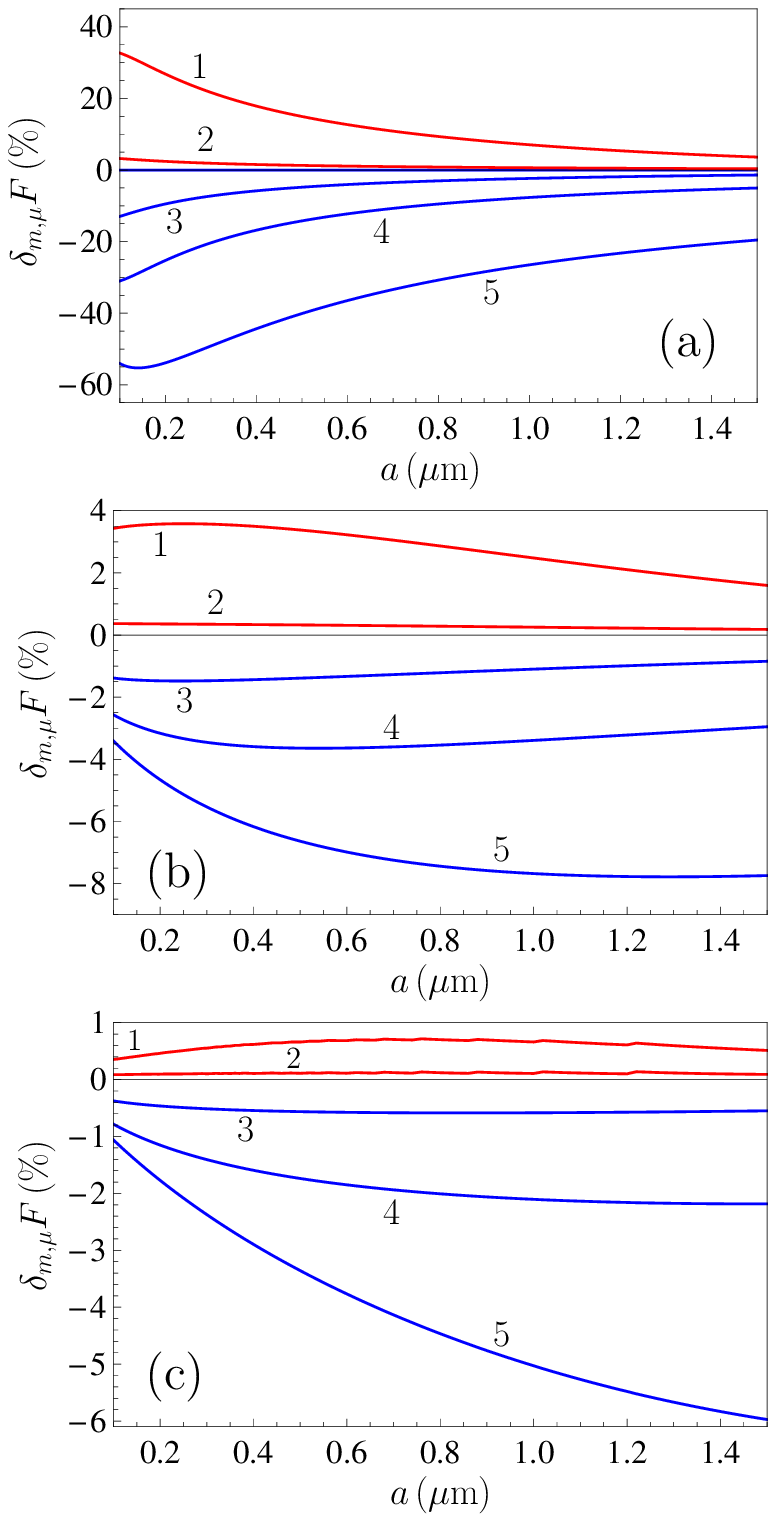}
}
\vspace*{-12cm}
\caption{\label{fg1}
The relative differences
of the Casimir force for real and pristine graphene  between
(a) a Au sphere  and a graphene sheet,
(b) a Au sphere  and a graphene-coated SiO${}_2$ plate  and
(c) a Au sphere  and  a graphene-coated Si plate are shown
as functions of separation at $T=300\,$K.
The lines 1, 2, 3, 4, and 5 {correspond to the following five combinations of values of $m$ and $\mu$:}
$m=0,\>\mu=0.5\,$eV; $m=0,\>\mu=0.1\,$eV;
$mc^2=0.1\,\mbox{eV},\>\mu=0$;
$mc^2=0.15\,\mbox{eV},\>\mu=0$;
and $mc^2=0.2\,\mbox{eV},\>\mu=0$, respectively.
}
\end{figure}
\begin{figure}[b]
\vspace*{-1cm}
\centerline{\hspace*{2.5cm}
\includegraphics{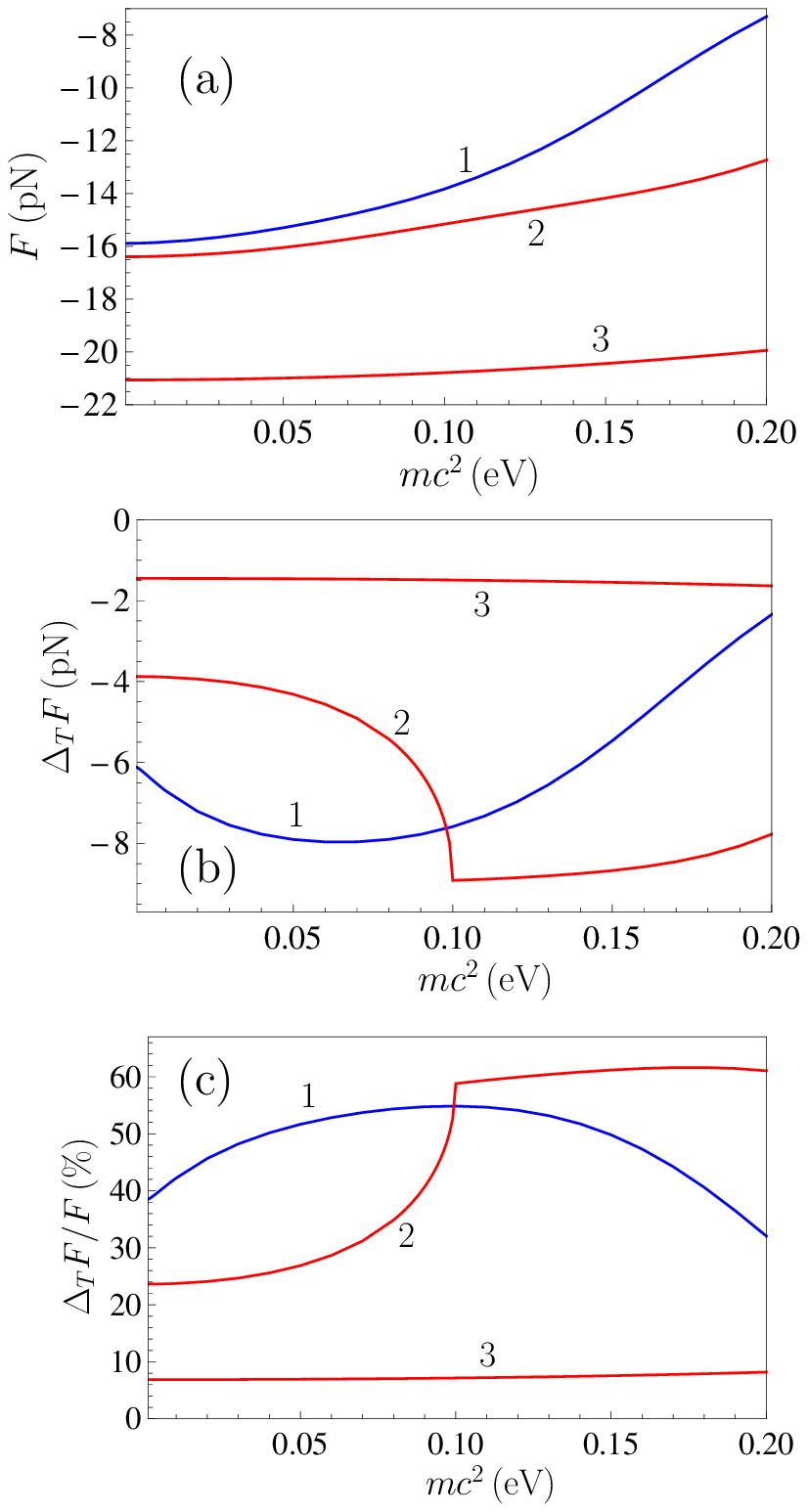}
}
\vspace*{-12cm}
\caption{\label{fg2}
(a) The Casimir force, (b) {its} thermal correction  and
(c) the fractional weight of the thermal correction in the Casimir force
for a Au sphere interacting with a graphene sheet
{for} $a=0.1\,\mu$m, $T=300\,$K
are shown as functions of the mass-gap parameter.
The lines 1, 2, and 3 are for $\mu=0$, 0.1, and 0.5\,eV,
respectively.
}
\end{figure}
\begin{figure}[b]
\vspace*{-1cm}
\centerline{\hspace*{2.5cm}
\includegraphics{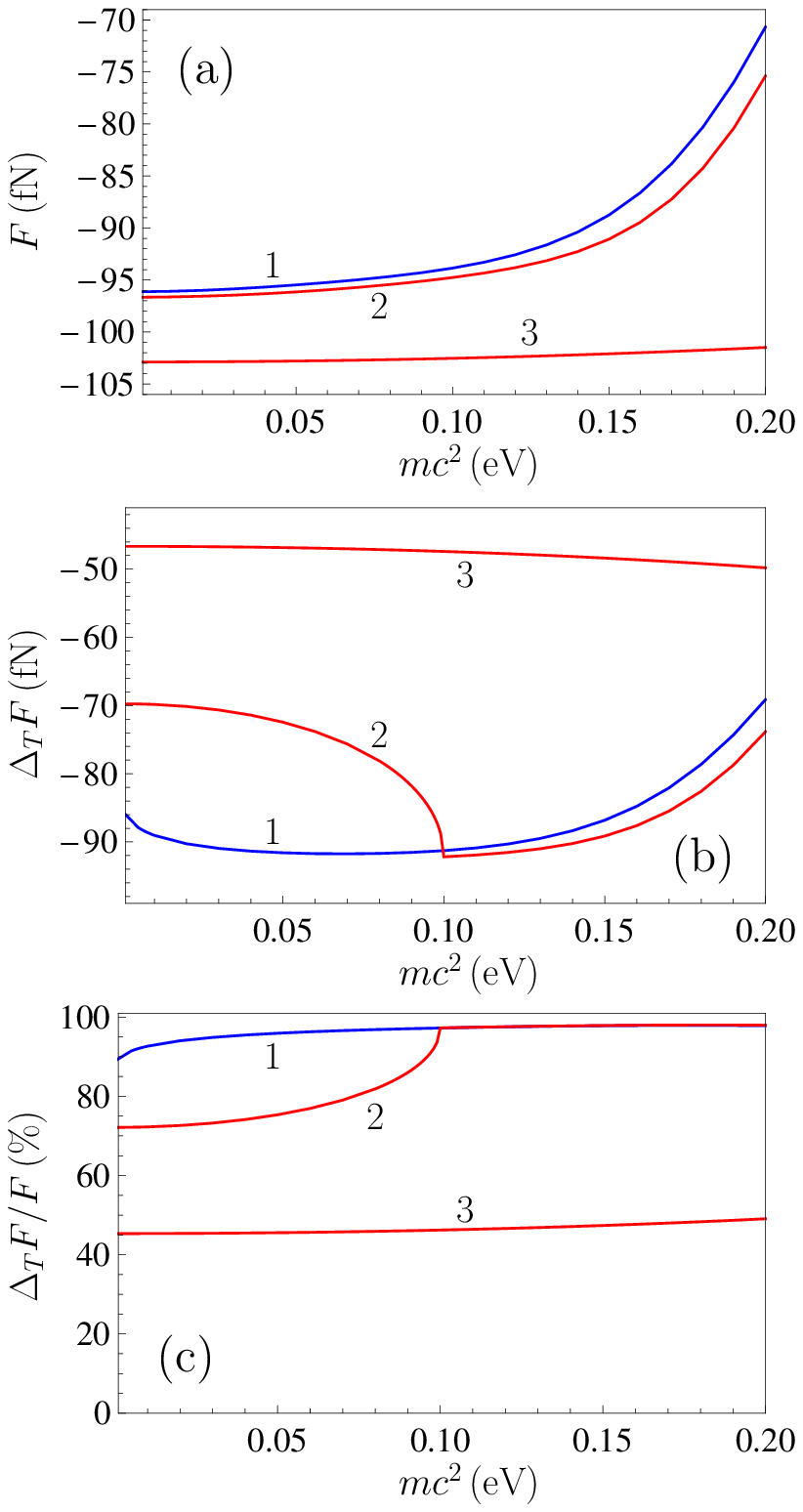}
}
\vspace*{-12cm}
\caption{\label{fg3}
(a) The Casimir force, (b) {its} thermal correction  and
(c) the fractional weight of the thermal correction in the Casimir force
for a Au sphere interacting with a graphene sheet
 {for} $a=1\,\mu$m, $T=300\,$K are shown
 as functions of the mass-gap parameter.
The lines 1, 2, and 3 are for $\mu=0$, 0.1, and 0.5\,eV,
respectively.
}
\end{figure}
\begin{figure}[b]
\vspace*{-1cm}
\centerline{\hspace*{2.5cm}
\includegraphics{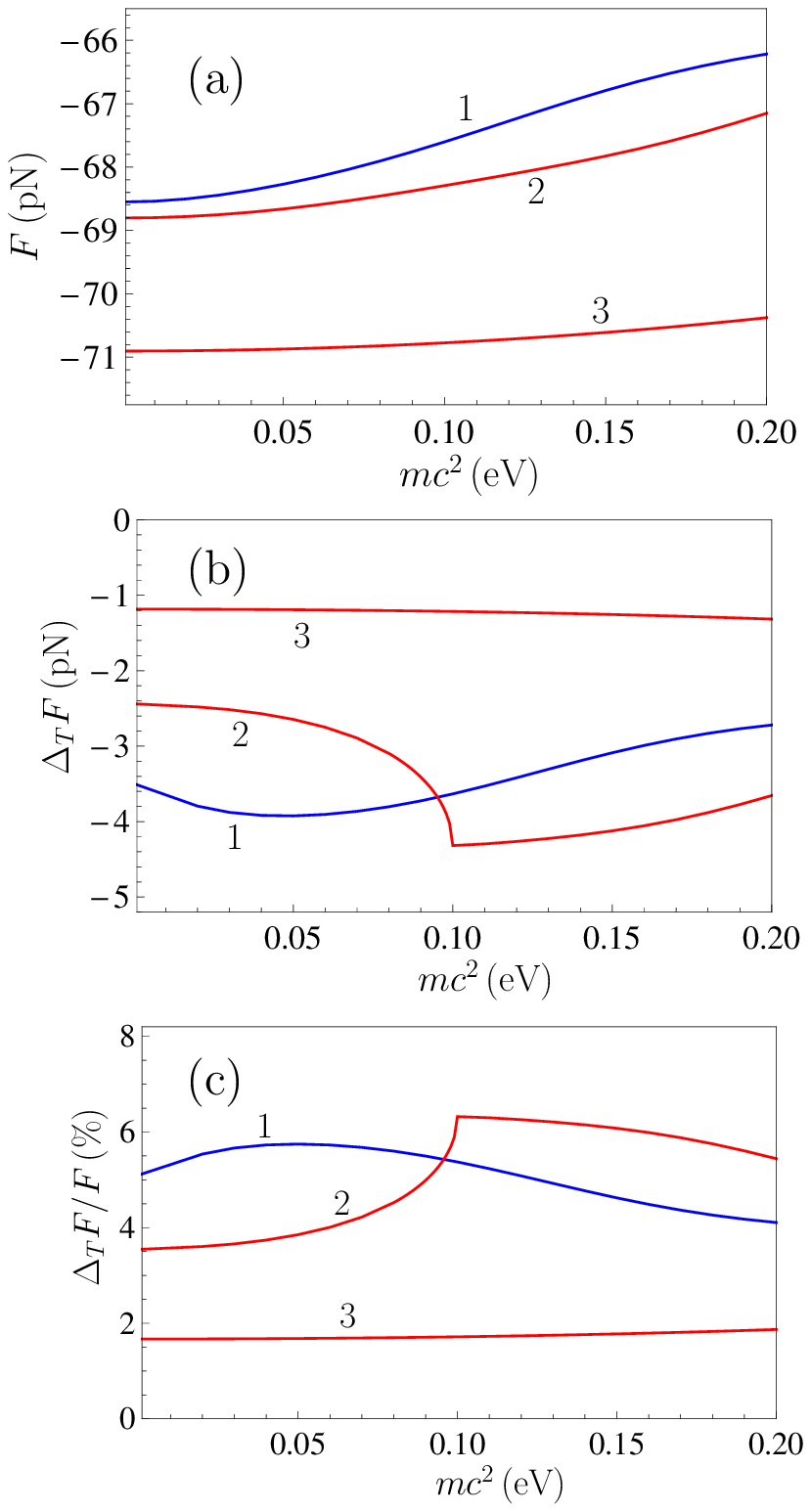}
}
\vspace*{-12cm}
\caption{\label{fg4}
(a) The Casimir force, (b) {its} thermal correction  and
(c) the fractional weight of the thermal correction in the Casimir force
for a Au sphere interacting with a graphene-coated SiO${}_2$ plate
{for} $a=0.1\,\mu$m, $T=300\,$K
are shown as functions of the mass-gap parameter.
The lines 1, 2, and 3 are for $\mu=0$, 0.1, and 0.5\,eV,
respectively.
}
\end{figure}
\begin{figure}[b]
\vspace*{-1cm}
\centerline{\hspace*{2.5cm}
\includegraphics{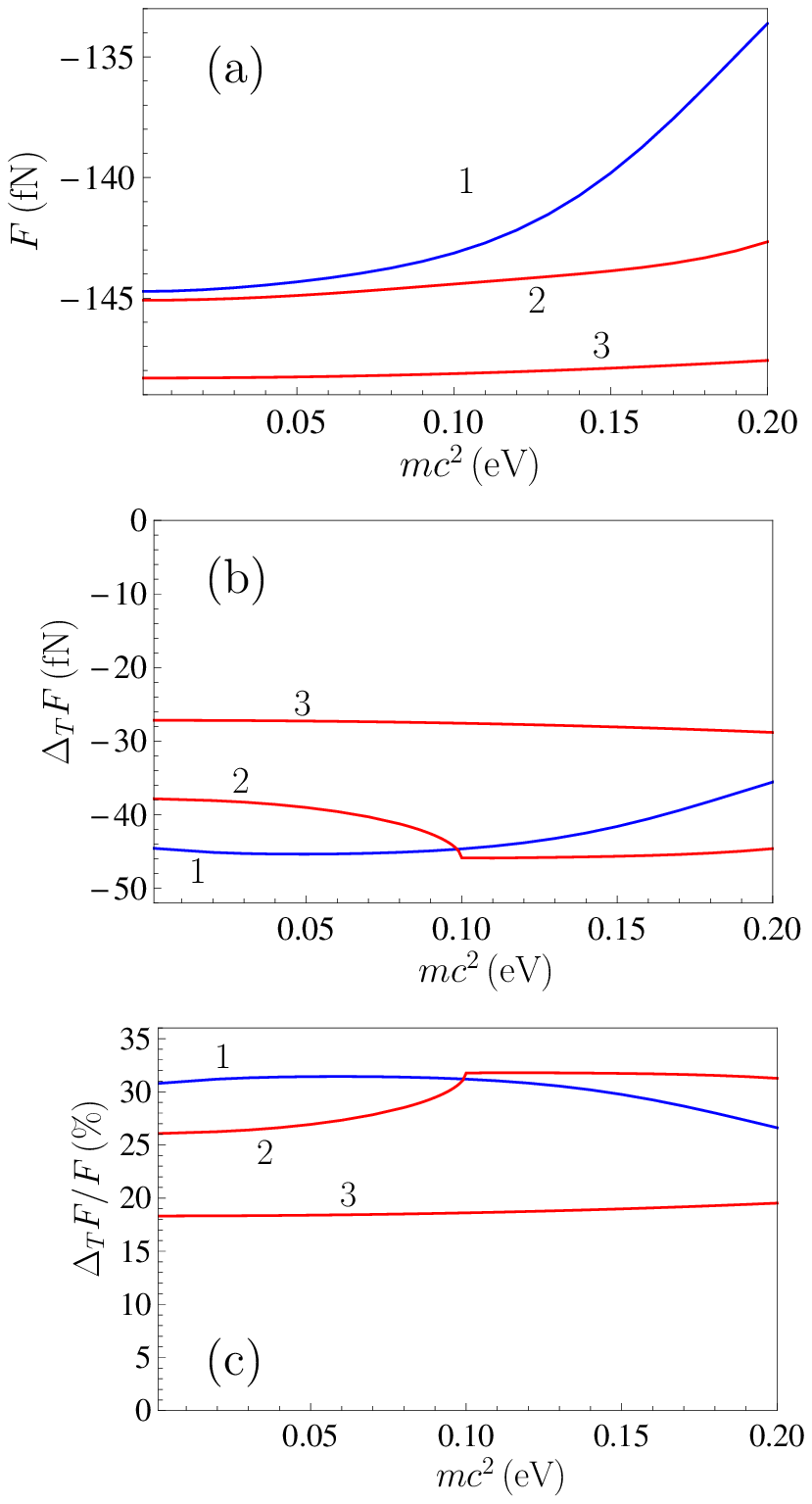}
}
\vspace*{-12cm}
\caption{\label{fg5}
(a) The Casimir force, (b) {its} thermal correction  and
(c) the fractional weight of the thermal correction in the Casimir force
for a Au sphere interacting with a graphene-coated SiO${}_2$ plate
 {for} $a=1\,\mu$m, $T=300\,$K are shown
 as functions of the mass-gap parameter.
The lines 1, 2, and 3 are for $\mu=0$, 0.1, and 0.5\,eV,
respectively.
}
\end{figure}
\begin{figure}[b]
\vspace*{-1cm}
\centerline{\hspace*{2.5cm}
\includegraphics{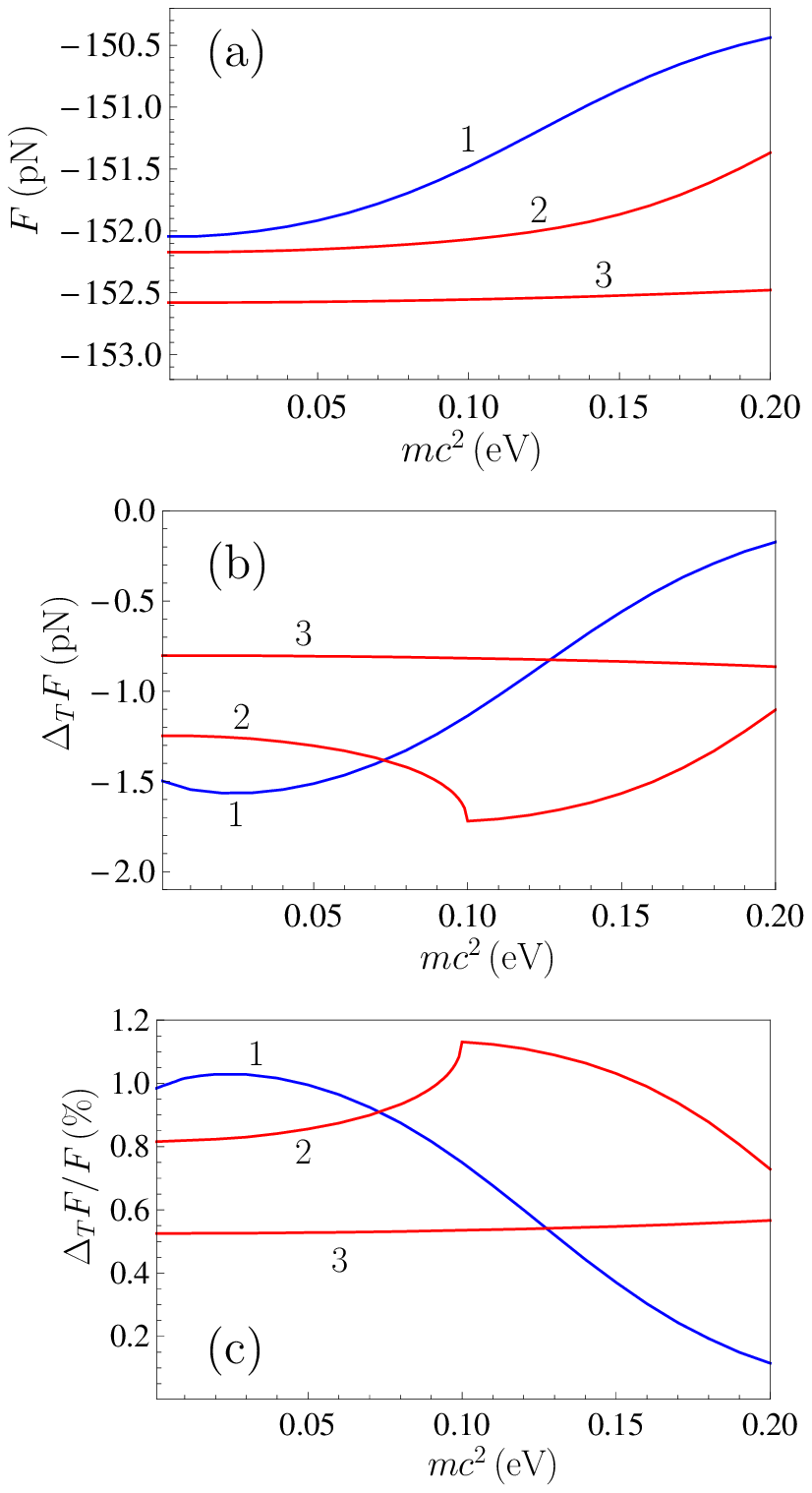}
}
\vspace*{-12cm}
\caption{\label{fg6}
(a) The Casimir force, (b) {its} thermal correction  and
(c) the fractional weight of the thermal correction in the Casimir force
for a Au sphere interacting with a graphene-coated Si plate
{for} $a=0.1\,\mu$m, $T=300\,$K
are shown as functions of the mass-gap parameter.
The lines 1, 2, and 3 are for $\mu=0$, 0.1, and 0.5\,eV,
respectively.
}
\end{figure}
\begin{figure}[b]
\vspace*{-1cm}
\centerline{\hspace*{2.5cm}
\includegraphics{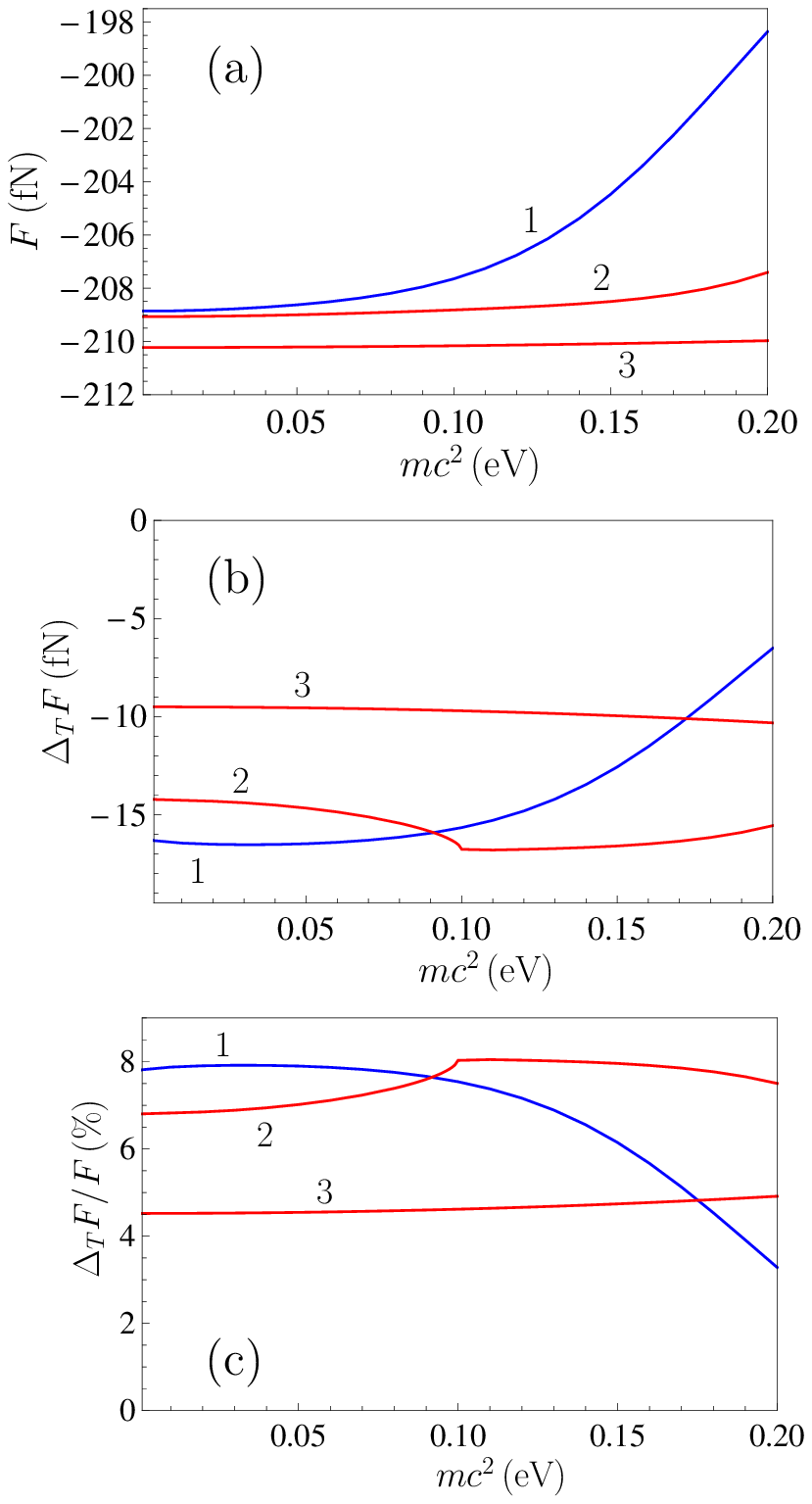}
}
\vspace*{-12cm}
\caption{\label{fg7}
(a) The Casimir force, (b) {its} thermal correction  and
(c) the fractional weight of the thermal correction in the Casimir force
for a Au sphere interacting with a graphene-coated Si plate
 {for} $a=1\,\mu$m, $T=300\,$K are shown
 as functions of the mass-gap parameter.
The lines 1, 2, and 3 are for $\mu=0$, 0.1, and 0.5\,eV,
respectively.
}
\end{figure}


\begin{thebibliography}{99}
\bibitem{1}
P.~J.~F.~Harris,
{\it Carbon Nanotubes and Related Structures: New Materials
for the Twenty-First Century}
(Cambridge University Press, New York, 1999).
\bibitem{2}
{\it Physics of Graphene}, ed. H.\ Aoki and M.\ S.\ Dresselhaus
(Springer, Cham, 2014).
\bibitem{3}
M.~I.~Katsnelson,
{\it Graphene: Carbon in Two Dimensions}
(Cambridge University Press, Cambridge, 2012).
\bibitem{4}
M.~O.~Goerbig,
Rev. Mod. Phys. {\bf 83}, 1193 (2011).
\bibitem{5}
M.~I.~Katsnelson, K.~S.~Novoselov, and A.~K.~Geim, Nature Phys. {\bf 2}, 620
(2006).
\bibitem{6}
D.~Allor, T.~D.~Cohen, and D.~A.~McGady,
Phys. Rev. D {\bf 78}, 096009 (2008).
\bibitem{7}
C.~G.~Beneventano, P.~Giacconi, E.~M.~Santangelo,
and R.~Soldati,
J. Phys. A {\bf 42}, 275401 (2009).
\bibitem{8}
G.~L.~Klimchitskaya
and V.~M.~Mostepanenko,
{Phys. Rev.} D {\bf 87}, 125011 (2013).
\bibitem{9}
I.~Akal, R.~Egger, C.~M\"{u}ller, and S.\ Villarba-Ch\'{a}vez,
{Phys. Rev.} D {\bf 93}, 116006 (2016).
\bibitem{10}
A.~W.~W.~Ludwig, M.~P.~A.~Fisher, R.~Shankar, and G.\ Grinstein,
Phys. Rev. B {\bf 50}, 7526 (1994).
\bibitem{11}
T.~Ando, Y.~Zheng, and H.~Suzuura,
J. Phys. Soc. Jpn. {\bf 71}, 1318 (2002).
\bibitem{12}
L.~A.~Falkovsky and A.~A.~Varlamov,
Eur. Phys. J. B {\bf 56}, 281 (2007).
\bibitem{13}
T.~Stauber, N.~M.~R.~Peres, and A.~K.~Geim,
Phys. Rev. B {\bf 78}, 085432 (2008).
\bibitem{14}
M.~Lewkowicz and B.~Rosenstein,
Phys. Rev. Lett. {\bf 102}, 106802 (2009).
\bibitem{15}
T.~Stauber,
J. Phys.: Condens. Matter {\bf 26}, 123201 (2014).
\bibitem{16}
M.~Merano,
Phys. Rev. A {\bf 93}, 013832 (2016).
\bibitem{17}
G.~L.~Klimchitskaya
and V.~M.~Mostepanenko,
{Phys. Rev.} B {\bf 93}, 245419 (2016).
\bibitem{18}
G.~L.~Klimchitskaya
and V.~M.~Mostepanenko,
{Phys. Rev.} B {\bf 94}, 195405 (2016).
\bibitem{19}
L.~A.~Falkovsky and S.~S.~Pershoguba,
Phys. Rev. B {\bf 76}, 153410 (2007).
\bibitem{20}
R.~R.~Nair, P.~Blake, A.~N.~Grigorenko, K.~S.~Novoselov, T.~J.~Booth, T.~Stauber,
N.~M.~R.~Peres, and A.~K.~Geim, Science {\bf 320}, 1308 (2008).
\bibitem{21}
F.~H.~L.~Koppens, D.~E.~Chang, and F.~J.~Garc\'{i}a de Abajo, Nano Lett. {\bf 11},
3370 (2011).
\bibitem{22}
M.~Bordag, G.~L.~Klimchitskaya, V.~M.~Mostepanenko, and V.~M.~Petrov,
Phys. Rev. D {\bf 91}, 045037 (2015); {\bf 93}, 089907(E) (2016).
\bibitem{23}
G.~L.~Klimchitskaya, C.~C.~Korikov, and V.~M.~Petrov,
Phys. Rev. B {\bf 92}, 125419 (2015); {\bf 93}, 159906(E) (2016).
\bibitem{24}
G.~L.~Klimchitskaya
and V.~M.~Mostepanenko,
{Phys. Rev.} A {\bf 93}, 052106 (2016).
\bibitem{24a}
G.~L.~Klimchitskaya
and V.~M.~Mostepanenko,
{Phys. Rev.} B {\bf 95}, 035425 (2017).
\bibitem{25}
M.~Bordag, G.~L.~Klimchitskaya, U.\ Mohideen, and
V.\ M.\ Mostepanenko, {\it Advances in the Casimir Effect}
(Oxford University Press, Oxford, 2015).
\bibitem{26}
E.~M.~Lifshitz and L.~P.~Pitaevskii,
{\it Statistical Physics}, Part II
(Pergamon, Oxford, 1980).
\bibitem{27}
G.~Barton,
J. Phys. A: Math. Gen. {\bf 38}, 2997 (2005).
\bibitem{28}
M.~Bordag,
J. Phys. A: Math. Gen. {\bf 39}, 6173 (2006).
\bibitem{29}
M.~Bordag, B.~Geyer, G.~L.~Klimchitskaya,
and V.~M.~Mostepanenko,
{Phys. Rev.} B {\bf 74}, 205431 (2006).
\bibitem{30}
M.~Bordag,
Phys. Rev. D {\bf 75}, 065003 (2007).
\bibitem{31}
E.~V.~Blagov, G.~L.~Klimchitskaya, and V.~M.~Mostepanenko,
{Phys. Rev. B} {\bf 75}, 235413 (2007).
\bibitem{32}
M.~Bordag and N.~Khusnutdinov,
Phys. Rev. D {\bf 77}, 085026 (2008).
\bibitem{33}
N.~R.~Khusnutdinov,
Phys. Rev. B {\bf 83}, 115454 (2011).
\bibitem{34}
G.~L.~Klimchitskaya
and V.~M.~Mostepanenko,
{Phys. Rev.} B {\bf 91}, 045412 (2015).
\bibitem{35}
A.~A.~Banishev, H.~Wen, J.~Xu, R.~K.~Kawakami,
G.~L.~Klimchitskaya, V.~M.~Mostepanenko, and U.~Mohideen,
Phys. Rev. B {\bf 87}, 205433 (2013).
\bibitem{36}
A.~H.~Castro Neto, F.~Guinea, N.~M.~R.~Peres, K.~S.~Novoselov,
and A.~K.~Geim,
Rev. Mod. Phys. {\bf 81}, 109 (2009).
\bibitem{37a}
N.~M.~R.~Peres,
Rev. Mod. Phys. {\bf 82}, 2673 (2010).
\bibitem{37}
J.~F.~Dobson, A.~White, and A.~Rubio,
{Phys. Rev. Lett.} {\bf 96}, 073201 (2006).
\bibitem{38}
G.~G\'{o}mez-Santos,
Phys. Rev. B {\bf 80}, 245424 (2009).
\bibitem{39}
D.~Drosdoff and L.~M.~Woods,
Phys. Rev. B {\bf 82}, 155459 (2010).
\bibitem{40}
D.~Drosdoff and L.~M.~Woods,
Phys. Rev. A {\bf 84}, 062501 (2011).
\bibitem{41}
Bo~E.~Sernelius,
Europhys. Lett. {\bf 95}, 57003 (2011).
\bibitem{42a}
T.~E.~Judd, R.~G.~Scott, A.~M.~Martin, B.\ Kaczmarek,
and T.\ M.\ Fromhold,
New J. Phys. {\bf 13}, 083020 (2011).
\bibitem{42}
J.~Sarabadani, A.~Naji, R.~Asgari, and R.~Podgornik,
Phys. Rev. B {\bf 84}, 155407 (2011);
Phys. Rev. B {\bf 87}, 239905(E) (2013).
\bibitem {43}
D.~Drosdoff, A.~D.~Phan, L.~M.~Woods, I.\ V.\ Bondarev,
and J.\ F.\ Dobson,
Eur. Phys. J. B {\bf 85}, 365 (2012).
\bibitem{44}
Bo~E.~Sernelius,
{Phys. Rev.} B {\bf 85}, 195427 (2012).
\bibitem{45}
A.~D.~Phan, L.~M.~Woods, D.~Drosdoff,
I.\ V.\ Bondarev, and N.\ A.\ Viet,
Appl. Phys. Lett. {\bf 101}, 113118 (2012).
\bibitem{46}
A.~D.~Phan, N.\ A.\ Viet, N.\ A.\ Poklonski, L.~M.~Woods,
 and  C.\ H.\ Le,
Phys. Rev. B {\bf 86}, 155419 (2012).
\bibitem{47a}
W.-K.~Tse and A.~H.~Macdonald,
Phys. Rev. Lett. {\bf 109}, 236806 (2012).
\bibitem{47b}
T.~Cysne, W.~J.~M.~Kort-Kamp, D.~Oliver, F.\ A.\ Pinheiro,
F.\ S.\ S.\ Rosa, and C.\ Farina,
Phys. Rev. A {\bf 90}, 052511 (2014).
\bibitem {47c}
V.\ B.\ Svetovoy and G.\ Palasantzas,
Phys. Rev. Applied {\bf 2}, 034006 (2014).
\bibitem{47}
N.~Knusnutdinov, R.~Kashapov,
 and L.~M.~Woods,
 Phys. Rev. A {\bf 94}, 012513 (2016).
\bibitem{48a}
T.~P.~Cysne, T.~G.~Rappoport, A.~Ferreira, J.\ M.\ Viana Parente Lopes,
and N.\ R.\ M.\ Peres,
Phys. Rev. B {\bf 94}, 235405 (2016).
\bibitem{48}
L.~M.~Woods, D.~A.~R.~Dalvit, A.~Tkatchenko, P.\ Rodrigues-Lopez,
A.\ W.\ Rodrigues, and R.\ Podgornik,
Rev. Mod. Phys. {\bf 88}, 045003 (2016).
\bibitem{49}
A.~I.~Akhiezer and V.~B.~Berestetskii, {\it Quantum Electrodynamics} (Interscience,
New York, 1965).
\bibitem{50}
S.~S.~Schweber, {\it An Introduction to Relativistic Quantum Field Theory}
(Dover, New York, 2005).
\bibitem{51}
M.~Bordag, I.~V.~Fialkovsky, D.~M.~Gitman, and
D.~V.~Vassilevich,
{Phys. Rev. B} {\bf 80}, 245406 (2009).
\bibitem{52}
I.~V.~Fialkovsky, V.~N.~Marachevsky, and
D.~V.~Vassilevich,
{Phys. Rev. B} {\bf 84}, 035446 (2011).
\bibitem{53}
M.~Bordag, G.~L.~Klimchitskaya, and
V.\ M.\ Mostepanenko,
Phys. Rev. B {\bf 86}, 165429 (2012).
\bibitem{54}
M.~Chaichian, G.~L.~Klimchitskaya, V.\ M.\ Mostepanenko,
and A.~Tureanu,
Phys. Rev. A {\bf 86}, 012515 (2012).
\bibitem{55}
G.~L.~Klimchitskaya
and V.~M.~Mostepanenko,
{Phys. Rev.} B {\bf 87}, 075439 (2013).
\bibitem{56}
B.~Arora, H.~Kaur, and B.~K.~Sahoo,
J. Phys. B {\bf 47}, 155002 (2014).
\bibitem{57}
K.~Kaur, J.~Kaur, B.~Arora,  and B.~K.~Sahoo,
Phys. Rev. B {\bf 90}, 245405 (2014).
\bibitem{58}
G.~L.~Klimchitskaya
and V.~M.~Mostepanenko,
{Phys. Rev.} A {\bf 89}, 012516 (2014).
\bibitem{59}
G.~L.~Klimchitskaya
and V.~M.~Mostepanenko,
{Phys. Rev.} B {\bf 89}, 035407 (2014).
\bibitem{60}
G.~L.~Klimchitskaya
and V.~M.~Mostepanenko,
{Phys. Rev.} A {\bf 89}, 052512 (2014).
\bibitem{61}
G.~L.~Klimchitskaya
and V.~M.~Mostepanenko,
{Phys. Rev.} A {\bf 89}, 062508 (2014).
\bibitem{62}
G.~L.~Klimchitskaya, V.~M.~Mostepanenko, and
Bo~E.~Sernelius,
Phys. Rev. B {\bf 89}, 125407 (2014).
\bibitem{63}
G.~L.~Klimchitskaya, U.~Mohideen, and V.~M.~Mostepanenko,
Phys. Rev. B {\bf 89}, 115419 (2014).
\bibitem{64}
G.~L.~Klimchitskaya
and V.~M.~Mostepanenko,
{Phys. Rev.} B {\bf 91}, 174501 (2015).
\bibitem{64a}
G.~L.~Klimchitskaya,
{Int. J. Mod. Phys.} A {\bf 31}, 1641026 (2016).
\bibitem{65}
V.~B.~Bezerra, G.~L.~Klimchitskaya,
V.~M.~Mostepanenko, and C.\ Romero,
{Phys. Rev.} A {\bf 94}, 042501 (2016).
\bibitem{66}
M.~Bordag, I.~Fialkovskiy, and D.~Vassilevich,
Phys. Rev. B {\bf 93}, 075414 (2016);
{\bf 95}, 119905(E) (2017).
\bibitem{67}
G.~Bimonte, G.~L.~Klimchitskaya, and V.~M.~Mostepanenko,
Phys. Rev. A {\bf 96}, 012517 (2017).
\bibitem{68}
C.~D.~Fosco, F.~C.~Lombardo, and F.~D.~Mazzitelli,
Phys. Rev. D {\bf 84}, 105031 (2011).
\bibitem{69}
G.~Bimonte, T.~Emig, R.~L.~Jaffe, and M.~Kardar,
Europhys. Lett. {\bf 97}, 50001 (2012).
\bibitem{70}
G.~Bimonte, T.~Emig, and M.~Kardar,
Appl. Phys. Lett. {\bf 100}, 074110 (2012).
\bibitem{71}
L.~P.~Teo,
Phys. Rev. D {\bf 88}, 045019 (2013).
\bibitem{71a}
M.~Hartmann, G.-L.~Ingold, and P.~A.~Maia Neto,
Phys. Rev. Lett. {\bf 119}, 043901 (2017).
\bibitem{71b}
G.~Bimonte,
Europhys. Lett. {\bf 118}, 20002 (2017).
\bibitem{72}
A.~Schmitt,
{\it Dense Matter in Compact Stars: A Pedagogical Introduction}
(Springer, Berlin, 2010).
\bibitem {73}
G.~L.~Klimchitskaya, U. Mohideen, and V.\ M.\ Mostepanenko,
 Rev. Mod. Phys. {\bf 81}, 1827 (2009).
\bibitem {74}
{\it Handbook of Optical Constants of Solids},
ed. E.~D.~Palik (Academic, New York, 1985).
\bibitem{75}
R.~S.~Decca,  E.~Fischbach, G.~L.~Klimchitskaya, D.~E.~Krause,
D.~L\'opez, and V.~M.~Mostepanenko,
Phys. Rev. D {\bf 68}, 116003 (2003).
\bibitem{76}
R.{\ }S. Decca, D. L\'opez, E. Fischbach, G.{\ }L. Klimchitskaya,
D.{\ }E. Krause, and V.{\ }M.\ Mostepanenko, Ann. Phys. (N.Y.) {\bf 318},
37 (2005).
\bibitem{77}
R.~S.~Decca, D.~L\'opez, E.~Fischbach, G.~L.~Klimchitskaya,
D.~E.~Krause, and V.~M.~Mostepanenko,
Phys. Rev. D {\bf 75}, 077101 (2007).
\bibitem{78}
R.~S.~Decca, D.~L\'opez, E.~Fischbach, G.~L.~Klimchitskaya,
D.~E.~Krause, and V.~M.~Mostepanenko,
Eur. Phys. J. C {\bf 51}, 963 (2007).
\bibitem{79}
C.-C.~Chang, A.~A.~Banishev, R.~Castillo-Garza,
G.~L.~Klimchitskaya, V.\ M.\ Mostepanenko, and U.\ Mohideen,
Phys. Rev. B {\bf 85}, 165443 (2012).
\bibitem{80}
A.~A.~Banishev,
G.~L.~Klimchitskaya, V.\ M.\ Mostepanenko, and U.\ Mohideen,
Phys. Rev. Lett. {\bf 110}, 137401 (2013).
\bibitem{81}
A.~A.~Banishev,
G.~L.~Klimchitskaya, V.\ M.\ Mostepanenko, and U.\ Mohideen,
Phys. Rev. B {\bf 88}, 155410 (2013).
\bibitem{82}
G.~Bimonte, D.~L\'{o}pez, and R.{\ }S.\ Decca,
Phys. Rev. B {\bf 93}, 184434 (2016).
\bibitem{83}
P.~K.~Pyatkovskiy,
J. Phys.: Condens. Matter {\bf 21}, 025506 (2009).
\bibitem{84}
V.~P.~Gusynin, S.~G.~Sharapov, and J.~P.~Carbotte,
New J. Phys. {\bf 11}, 095013 (2009).
\bibitem{85}
S.~A.~Jafari,
J. Phys.: Condens. Matter {\bf 24}, 205802 (2012).
\bibitem{86}
E.~M.~Hajaj, O.~Shtempluk, V.~Kochetkov, A.\ Razin, and Y.\ E.\ Yaish,
Phys. Rev. B {\bf 88}, 045128 (2013).
\bibitem{86a}
K.~F.~Mak, M.~Y.~Sfeir, Y.~Wu, C.~H.~Lui, J.\ A.\ Misewich,
and T.\ F.\ Heinz,
Phys. Rev. Lett. {\bf 101}, 196405 (2008).
\bibitem{87}
D.~B.~Hough and L.~R.~White,
Adv. Coll. Interface Sci. {\bf 14}, 3 (1980).
\bibitem{87a}
L.~Bergstr\"{o}m,
Adv. Coll. Interface Sci. {\bf 70}, 125 (1997).
\bibitem{88}
F.~Chen, G.{\ }L.\ Klimchitskaya, V.{\ }M.\ Mostepanenko, and U.\ Mohideen,
Phys. Rev. B {\bf 76}, 035338 (2007).
\bibitem{89}
R.~S.~Decca, D.~L\'opez, H.~B.~Chan, E.~Fischbach,
 D.~E.~Krause,  and C.~R.~Jamell,
Phys. Rev. Lett. {\bf 94}, 240401 (2005).
\bibitem{90}
Y.-J.~Chen, W.~K.~Tham,
 D.~E.~Krause, D.~L\'opez, E.~Fischbach,
 and R.~S.~Decca,
{Phys. Rev. Lett.} {\bf 116}, 221102  (2016).
\bibitem{91}
G. Bimonte, Phys. Rev. Lett. {\bf 112}, 240401 (2014).
\bibitem{92}
G. Bimonte, Phys. Rev. Lett. {\bf 113}, 240405 (2014).
\bibitem{93}
G. Bimonte, Phys. Rev. B {\bf 91}, 205443 (2015).
\bibitem{94}
G.~Bimonte, G.~L.~Klimchitskaya, and V.~M.~Mostepanenko,
Phys. Rev. A  {\bf 95}, 052508 (2017).


\end{thebibliography}
\end{document}